\documentclass[11pt]{article}
\usepackage{amsfonts}
\usepackage{amssymb}
\usepackage{amscd}

\def\a{\alpha}
\def\b{\beta}
\def\g{\gamma}
\def\d{\delta}
\def\e{\epsilon}
\def\l{\lambda}
\def\o{\omega}
\def\p{\partial}
\def\ve{\varepsilon}
\def\vp{\varphi}
\def\s{\sigma}
\def\th{\theta}

\newcommand{\cg}{{\mathcal G}}
\newcommand{\cl}{{\mathcal L}}
\newcommand{\cm}{{\mathcal M}}
\newcommand{\cn}{{\mathcal N}}
\newcommand{\co}{{\mathcal O}}
\newcommand{\cp}{{\mathcal P}}
\newcommand{\mr}{{\mathcal R}}
\newcommand{\ct}{{\mathcal T}}
\newcommand{\cz}{{\mathcal Z}}
\newcommand{\U}{{\mathcal U}}

\newcommand{\C}{\mathbb C}
\newcommand{\R}{\mathbb R}
\newcommand{\T}{\mathbb T}
\newcommand{\Z}{\mathbb Z}

\newcommand{\sdiff}{{\mathop{\mbox{sdiff}}\nolimits\,}}
\newcommand{\SDiff}{{\mathop{\mbox{SDiff}}\nolimits\,}}
\newcommand{\LSDiff}{{\mathop{\mbox{LSDiff}}\nolimits\,}}
\newcommand{\Vect}{{\mathop{\mbox{Vect}}\nolimits\,}}

\def\D{\Delta}
\def\h{\eta}
\def\m{\mu}
\def\n{\nu}
\def\c{\chi}
\def\j{\psi}

\def\ba{{\bar{A}}}
\def\bb{{\bar{B}}}
\def\bw{{\bar{w}}}
\def\bz{{\bar{z}}}
\def\by{{\bar{y}}}
\def\bc{{\bar{\c}}}
\def\bj{{\bar{\j}}}

\def\tj{{\tilde{j}}}
\def\tl{{\tilde{\ell}}}
\def\tm{{\tilde{m}}}

\def\pa{\partial}
\def\dt#1{{\buildrel {\hbox{\LARGE .}} \over {#1}}}  
\def\ad{{\dt{\alpha}}}
\def\bd{{\dt{\beta}}}
\def\pd{{\dt{+}}}
\def\md{{\dt{-}}}
\def\dvec#1{\buildrel \leftrightarrow \over #1}
\def\sfrac#1#2{{\textstyle\frac#1#2}}

\def\beq{\begin{equation}}
\def\eeq{\end{equation}}
\def\beqx{\begin{displaymath}} 
\def\eeqx{\end{displaymath}}
\def\beql{\arraycolsep .1em \begin{eqnarray}}
\def\eeql{\end{eqnarray}}

\def\gl#1{(\ref{#1})}

\def\theequation{\thesection.\arabic{equation}}

\catcode`@=11
\@addtoreset{equation}{section}
\@addtoreset{equation}{subsection}
\def\theequation{\ifnum\value{section}=0 \arabic{equation}\ignorespaces
\else \ifnum\value{section}=-1 A.\arabic{equation}\ignorespaces
\else \ifnum\value{subsection}=0 \thesection.\arabic{equation}\ignorespaces
\else \thesection.\arabic{subsection}.\arabic{equation}\ignorespaces
                           \fi
                      \fi
                 \fi}
\catcode`@=12

\def\be{\begin{equation}}
\def\ee{\end{equation}}
\def\bea{\begin{eqnarray}}
\def\eea{\end{eqnarray}}

\begin{document}
\begin{titlepage}
\begin{flushright}
hep-th/9912154\\
ITP--UH--23/99 \\
December, 1999
\end{flushright}

\vskip 2.0cm

\begin{center}
{\Large\bf Closed N=2 Strings: Picture-Changing,}

\vspace{0.2cm}

{\Large\bf Hidden Symmetries and SDG Hierarchy}

\vskip 1.5cm

{\Large \ Olaf Lechtenfeld}

\vskip 0.5cm

{\it Institut f\"ur Theoretische Physik, Universit\"at Hannover}\\
{\it Appelstra\ss{}e 2, 30167 Hannover, Germany}\\
{E-mail: lechtenf@itp.uni-hannover.de}

\vskip 0.5cm
{\large and}
\vskip 0.5cm

{\Large Alexander D. Popov}

\vskip 0.5cm

{\it Bogoliubov Laboratory of Theoretical Physics}\\
{\it JINR, 141980 Dubna, Moscow Region, Russia}\\
{E-mail: popov@thsun1.jinr.ru}

\end{center}
\vskip 1.5cm

\begin{abstract}

We study the action of picture-changing and spectral flow operators
on a ground ring of ghost number zero operators in the chiral BRST
cohomology of the closed $N{=}2$ string and describe an infinite set
of symmetry charges acting on physical states.
The transformations of physical string states are compared with symmetries
of self-dual gravity which is the effective field theory of the closed
$N{=}2$ string.
We derive all infinitesimal symmetries of the self-dual gravity equations
in $2{+}2$ dimensional spacetime and introduce an infinite hierarchy of
commuting flows on the moduli space of self-dual metrics.
The dependence on moduli parameters can be recovered by solving the
equations of the SDG hierarchy associated with an infinite set of abelian 
symmetries generated recursively from translations.
These non-local abelian symmetries are shown to coincide with the hidden
abelian string symmetries responsible for the vanishing of most scattering
amplitudes.
Therefore, $N{=}2$ string theory ``predicts'' not only self-dual gravity
but also the SDG hierarchy.

\end{abstract}

\vfill
\end{titlepage}

\newpage


\section{Introduction}

The pioneering work of Ooguri and Vafa~\cite{OV} revealed an intimate
connection between self-dual field theories and (classical) $N{=}2$
string theories, formulated in four spacetime dimensions.
In particular, four-dimensional manifolds with a metric $g$ of
ultrahyperbolic signature $(++-\,-)$ and a self-dual Riemann tensor
arise as exact (to all orders in $\a'$) classical background configurations
for the closed $N{=}2$ string.
Indeed, the only physical string degree of freedom in this case is
a massless scalar field whose (tree-level) dynamics takes the form of
Plebanski's first~\cite{OV} or second~\cite{LS} equation,
which both describe self-dual gravity (SDG)~\cite{Ple},
albeit in different gauges.
Although the absence of an infinite tower of massive excitations indicates
a sort of caricature of a string, this quality makes it amenable to exact
solutions, a fact quite rare in string theory.
Yet $N{=}2$ strings may not only serve as a testing ground for certain issues
in string theory in general but, being consistent quantum theories,
they can also serve as a guide in the quantization of self-dual gravity.

\medskip

The self-duality equations for the Riemann tensor can be considered
on {\it complex\/} four-manifolds $M^\C$ with holomorphic metric $g^\C$,
and in most papers on SDG just the complex case was considered.
In order to investigate the symmetries of the SDG equations,
one usually fixes some special form of a tetrad,
which corresponds to the choice of a gauge.
This choice breaks the invariance of the SDG equations
under the gauge group SDiff$(M^\C)$ of volume-preserving
diffeomorphisms of a manifold $M^\C$.  Using different gauges,
various symmetries of the SDG equations were uncovered
(see e.g.~\cite{Ta,Pa,PBR} and references therein).
It is fair to say, however, that the connections between these
symmetries have not yet been clarified.
Also missing is a discussion of the symmetry subalgebras
compatible with a real structure on $M^\C$,
i.e. symmetry algebras of the SDG equations on Riemannian
or Kleinian four-dimensional manifolds with signature
$(4,0)$ or $(2,2)$, respectively.
Again, the first steps in this direction have been made
by Ooguri and Vafa~\cite{OV}.

\medskip

We notice that the description of self-duality depends on the
orientation of the manifold $M$, and self-duality can be replaced
by anti-self-duality upon changing the orientation of $M$.
This paper will be concerned with the SDG equations.
We shall describe the SDG equations on Kleinian four-dimensional manifolds $M$
of signature (2,2) and discuss their integrability, hidden symmetries,
and hierarchies. No complete treatment of these problems yet exists.
We shall also discuss a connection between the group-theoretic and
the geometric twistor approaches to the symmetries of the SDG equations,
describe a general solution of the linearized SDG equations, and
therefore present {\it all\/} infinitesimal symmetries of these equations.

\medskip

If $N{=}2$ string theory ``predicts'' self-dual gravity,
its wealth of symmetries should be obtainable from the stringy description.
More precisely,
we expect the SDG hierarchy related to the abelian symmetries of the
SDG equations to be visible in $N{=}2$ closed string quantum mechanics.
Indeed, in an earlier paper with J\"unemann~\cite{JLP},
the authors have recently identified part of these hidden string symmetries
and have demonstrated that they are the cause of the vanishing of
almost all scattering amplitudes.
Quite surprisingly, the stringy root of such symmetries is technically the
somewhat obscure picture phenomenon~\cite{FMS} which is present whenever
covariant quantization meets worldsheet supersymmetry.
Global symmetries unbroken by the string background under consideration
may be classified with the help of BRST cohomology, and the latter
unexpectedly displays a picture dependence~\cite{JL} (see also~\cite{BZ}).
This connection hints at a geometrical interpretation of the picture
phenomenon of the closed $N{=}2$ string in terms of flows in the moduli
space of self-dual metrics.

\section{Review of the closed N=2 string}

{}From the worldsheet point of view, critical closed $N{=}2$ strings
in flat Kleinian space $\R^{2,2}$
are a theory of $N{=}(2,2)$ supergravity $(h,\c,A)$ in $1{+}1$ dimensions
coupled to two chiral $N{=}(2,2)$ massless matter multiplets $(y,\j)$.
The latter's components are complex scalars (the four string coordinates)
and $SO(1,1)$ Dirac spinors (their four NSR partners).
The $N{=}2$ string Lagrangian,
as first written down by Brink and Schwarz~\cite{BS}, reads
\bea
\cl\ &=&\ \sqrt{h}\,\Bigl\{
          \sfrac12 h^{mn}\pa_m \by^\ba \pa_n y^A
         +\sfrac{i}2 \bj^{-\ba} \g^{m} \dvec{D}_m \j^{+A}
         +A_m \bj^{-\ba} \g^{m} \j^{+A} \nonumber\\[.7ex]
      && +\,(\pa_m \by^\ba + \bj^{-\ba} \c^+_m)
          \bc^-_n \g^m \g^n \j^{+A}
         +\bj^{-\ba} \g^n \g^m \c^+_n
          (\pa_m y^A + \bc^-_m \j^{+A}) \Bigr\}\,\h_{\ba A}
\eea
where
$h_{mn}$ and $A_m$, with $m{=}0,1$, are the (real) worldsheet
metric and $U(1)$ gauge connection, respectively.
The worldsheet gravitino $\c_m$ as well as the matter fields
$y^A$ and $\j^A$ are complex valued, so that the spacetime index
$A,\ba=1,2$ runs over two values only. Complex conjugation reads
\be
(y^A)^*\ =\ \by^{\ba} \qquad{\rm but}\qquad
(\j^{+A})^*\ =\ \j^{-\ba} \quad{\rm and}\quad (\c^+_m)^*\ =\ \c^-_m \quad,
\ee
and $\h_{\ba A}={\rm diag}(+-)$ is the flat metric in~$\C^{1,1}$.
As usual, $\{\g^m\}$ are a set of $SO(1,1)$ worldsheet gamma matrices,
$\bj=\j^\dagger\g^0$, and
$D_m$ denotes the worldsheet gravitationally covariant derivative.

\medskip

This formulation entails the choice of a complex structure on Kleinian space.
A given complex structure breaks the global ``Lorentz'' invariance
of $\R^{2,2}$,
\be \label{break1}
{\rm Spin}(2,2)\ =\ SU(1,1) \times SU(1,1)' \
\longrightarrow\ U(1) \times SU(1,1)' \ \simeq\ U(1,1) \quad.
\ee
The moduli space of complex structures is the two-sheeted hyperboloid
$H^2=H_+^2\cup H_-^2$ with $H_\pm^2\simeq SU(1,1)/U(1)$.
It can be completed to $CP^1$ by sewing the two sheets together along
a circle,
\be \label{sew}
CP^1\ =\ H_+^2 \cup S^1 \cup H_-^2 \quad.
\ee

\medskip

Instead of using complex coordinates adapted to $SU(1,1)'$,
one may alternatively choose a basis appropriate for $SL(2,\R)'$
and employ a real notation for the string coordinates,
\be
y^1\ =\ x^1 + i x^2 \quad, \qquad y^2\ =\ x^3 + i x^4 \quad,
\ee
by expressing the real coordinates $x^\m$, $\m,\n,\ldots=1,2,3,4$, in
$SL(2,\R)\times SL(2,\R)'$ spinor notation,
\be
x^{\a\ad}\ =\ \s_\m^{\a\ad} x^\m\ =\
\left(\begin{array}{cc}
x^4{+}x^2 & x^1{-}x^3 \\ x^1{+}x^3& x^4{-}x^2
\end{array}\right) \quad,\qquad
\a\in\{+,-\} \; , \quad \ad\in\{\pd,\md\} \;,
\ee
with the help of chiral gamma matrices $\s_\mu$
appropriate for the spacetime metric $\h_{\mu\nu}={\rm diag}(++--)$.

\medskip

In the real formulation,
the tangent space at any point of $\R^{2,2}$ can be split to $\R^2\oplus\R^2$
which defines a real polarization or cotangent structure.
Such a polarization is characterized by a pair of null planes $\R^2$,
and the latter are determined by a real null two-form modulo scale or,
equivalently, by a real $SL(2,\R)$ spinor~$v$ modulo scale.
Indeed, each null vector~$(u_{\a\ad})$ factorizes into two real spinors,
$u_{\a\ad}=v_\a w_\ad$.
Choosing coordinates such that ${v_+\choose v_-}={t\choose0}$, it
becomes clear that a given null plane is stable under the action of
\be \label{borel}
B_+ \times SL(2,\R)' \quad, \qquad{\rm with} \quad B_+\ :=\
\Bigl\{ {a\ \ b \choose\ 0\ a^{-1}\!} \Bigm| a\in\R^*,\; b\in\R \Bigr\}\quad,
\ee
where $B_+$ acts on $v$ and $SL(2,\R)'$ on $w$.
The moduli space of cotangent structures thus becomes
\be
{\rm Spin}(2,2)/[B_+\times SL(2,\R)']\ \simeq\
SL(2,\R)/B_+ \ \simeq\ S^1
\ee
which in fact is just the $S^1$ in \gl{sew}.
However, it turns out~\cite{BL1} that the real spinor $v$ also encodes the
two string couplings,
\be
{v_+\choose v_-}\ =\ \kappa^{1/4}\,{\cos\sfrac\th2 \choose \sin\sfrac\th2}
\ee
with $\kappa\in\R^+$ being the gravitational coupling and $\th\in S^1$
the instanton angle.
Since $v$ (including scale) is inert only under the parabolic subgroup
of $B_+$ obtained by putting $a{=}1$,
the space of string couplings is that of nonzero real $SL(2,\R)$ spinors,
\be
\R^+\times S^1\ \simeq\ \R^2-\{0\}\ \simeq\ \C-\{0\}\
\ni\ \kappa^{1/4}\;e^{i\th/2} \quad.
\ee
Consequently, fixing the values of the string couplings amounts
to breaking the global ``Lorentz'' invariance of $\R^{2,2}$ in a way
different from \gl{break1},
\be \label{break2}
{\rm Spin}(2,2)\ =\ SL(2,\R) \times SL(2,\R)' \
\longrightarrow\ \R \times SL(2,\R)' \quad,
\ee
where $\R\simeq B_+(a{=}1)$ from eq.\gl{borel}.

\medskip

The $N{=}2$ supergravity multiplet defines a gravitini and a Maxwell
bundle over the worldsheet Riemann surface.
The topology of the total space is labeled by the Euler number~$\chi$
of the punctured Riemann surface and
the first Chern number (instanton number)~$M$ of the Maxwell bundle.
It is notationally convenient to replace the Euler number by the ``spin''
\be
J\ :=\ -2\chi\ =\ 2n-4+4(\# {\rm handles})\ \in\ 2\Z \quad.
\ee
The Lagrangian is to be integrated over the string worldsheet of a given
topology. The first-quantized string path integral for the $n$-point
function $A^{(n)}$ includes a sum over worldsheet topologies~$(J,M)$,
weighted with appropriate powers in the
string couplings~$(\kappa,e^{i\th})$:
\be
A^{(n)}\ =\ \sum_{J=2n-4}^{\infty} \kappa^{J/2}\,A^{(n)}_J\ =\
\sum_{J=2n-4}^{\infty}\sum_{M=-J}^{+J}\kappa^{J/2}\,e^{iM\th}\,A^{(n)}_{J,M}
\ee
where the instanton sum has a finite range because bundles with $|M|{>}J$
do not contribute.
The presence of Maxwell instantons breaks the explicit $U(1)$ factor in
\gl{break1} but the $SU(1,1)$ factor (and thus the whole ${\rm Spin}(2,2)$)
is fully restored if we let
$\kappa^{1/4}(e^{i\th/2},e^{-i\th/2})$ transform as an $SU(1,1)$ spinor.
The partial amplitudes $A^{(n)}_{J,M}$ are integrals over the metric,
gravitini, and Maxwell moduli spaces. The integrands may be obtained
as correlation functions of vertex operators in the $N{=}(2,2)$ superconformal
field theory on the worldsheet surface of fixed shape (moduli) and
topology.

\medskip

The vertex operators generate from the (first-quantized)
vacuum state the asymptotic string states in the
scattering amplitude under consideration.
They uniquely correspond to the physical states of the $N{=}2$ closed string
and carry their quantum numbers.
The physical subspace of the $N{=}2$ string Fock space in a covariant
quantization scheme turns out to be surprisingly small~\cite{Bi}:
Only the ground state $|k\rangle$ remains, a scalar on the
massless level, i.e. for center-of-mass momentum~$k^A$ with
$\bar{k}\cdot k:=\eta_{\ba A} \bar{k}^\ba k^A=0$.
The dynamics of this string ``excitation'' is described by a
massless scalar field,
\be
\Phi(y)\ =\ \int\!\!d^4k\;e^{-i(\bar{k}\cdot y+k\cdot\by)}\;\tilde{\Phi}(k)
\quad,
\ee
whose self-interactions are determined on-shell from the
(amputated tree-level) string scattering amplitudes,
\be
\langle \tilde{\Phi}(k_1)\,\tilde{\Phi}(k_2)\ldots\tilde{\Phi}(k_n)
\rangle^{\rm amp}_{{\rm tree},\th}\ =:\
A^{(n)}_{2n-4}(k_1,\ldots,k_n;\th)\ =:\
\d_{k_1{+}\ldots{+}k_n}\;\tilde{A}^{(n)}_{2n-4}(k_1,\ldots,k_n;\th)
\quad.
\ee
Interestingly, it has been shown~\cite{OV,Hipp}
that all tree-level $n$-point functions vanish on-shell,
except for the two- and three-point amplitudes,
\bea
\tilde{A}^{(2)}_0(k_1,k_2;\th) &=& 1 \quad, \\[1ex]
\tilde{A}^{(3)}_2(k_1,k_2,k_3;\th) &=& -\frac14 \Bigl[ \,
\e_{AB}\,k_1^A\,k_2^B\,e^{i\th}\ -\
\h_{A\bb}\,(k_1^A\,\bar{k}_2^\bb-\bar{k}_1^\bb\,k_2^A)\ -\
\e_{\ba\bb}\,\bar{k}_1^\ba\,\bar{k}_2^\bb\,e^{-i\th} \Bigr]^{\textstyle2}
\nonumber\\[1ex]
&=& \Bigl[ \, \e_{\ad\bd} \, \Bigl(
k_1^{+\ad}k_2^{+\bd} \cos^2\sfrac\th2 +
(k_1^{+\ad}k_2^{-\bd}+k_1^{-\ad}k_2^{+\bd}) \cos\sfrac\th2\sin\sfrac\th2 +
k_1^{-\ad}k_2^{-\bd} \sin^2\sfrac\th2 \Bigr) \Bigr]^{\textstyle2}
\nonumber\\[1ex]
&=& \Bigl[ \, \e_{\ad\bd} \, \Bigl(
k_1^{+\ad} \cos\sfrac\th2 + k_1^{-\ad} \sin\sfrac\th2 \Bigr) \Bigl(
k_2^{+\bd} \cos\sfrac\th2 + k_2^{-\bd} \sin\sfrac\th2 \Bigr)
\Bigr]^{\textstyle2} \quad,
\eea
with $\bar{k}_i\cdot k_j+\bar{k}_j\cdot k_i=0$ due to $\sum_n k_n=0$.
Note that $\tilde{A}^{(3)}_2$ is totally symmetric in all momenta.

\medskip

Since we argue that the string couplings $(\kappa,e^{i\th})$
can be changed at will by global ``Lorentz'' transformations,
it is admissible to make a convenient choice of Lorentz frame.
First, we may scale $\kappa\to1$ (i.e. put the constant dilaton to zero).
Second, the instanton angle $\th$ is at our disposal.
In the real notation, renaming $\Phi\to\Psi$,
one sees that taking $\th{=}0$ reduces the amplitude
to its $M{=}{+}J$ contribution~\cite{LS},
\be
\tilde{A}^{(3)}_2(k_1,k_2,k_3;\th{=}0)\ =\
\Bigl[ \e_{\ad\bd}\, k_1^{+\ad}\, k_2^{+\bd} \Bigr]^2 \quad,
\ee
which translates to a cubic interaction~\footnote{
The $SO(2,2)$ transformation properties of this interaction
become manifest when this term is rewritten as
$\sfrac{\kappa}6\,T^{(+)}_{\a\g\b\d}\,\e^{\ad\bd}\,\e^{\dt{\g}\dt{\d}}\;
\Psi\;\pa^\a_{\ad}\pa^\g_{\dt{\g}}\Psi\;\pa^\b_{\bd}\pa^\d_{\dt{\d}}\Psi$,
with a self-dual projector $T^{(+)}$ having nonzero components
$T^{(+)}_{++++}=1$ only.}
\be
\cl_{\rm int}\ =\ \sfrac{\kappa}6\,\e^{\ad\bd}\,\e^{\dt{\g}\dt{\d}}\;
\Psi\;\pa^+_{\ad}\pa^+_{\dt{\g}}\Psi\;\pa^+_{\bd}\pa^+_{\dt{\d}}\Psi \quad.
\ee
The resulting equation of motion reads
\be \label{ple2}
-\square\,\Psi\ +\ \sfrac{\kappa}2\,\e^{\ad\bd}\,\e^{\dt{\g}\dt{\d}}\;
\pa^+_{\ad}\pa^+_{\dt{\g}}\Psi\;\pa^+_{\bd}\pa^+_{\dt{\d}}\Psi\ =\ 0
\ee
and is known as Plebanski's {\it second\/} equation.
It describes the dynamics of the single-helicity ($h{=}{+}2$) graviton
in $2{+}2$ self-dual gravity.
More precisely, the self-dual Riemann tensor reduces to the
(0,2) Weyl tensor~$C_{\ad\bd\dt{\g}\dt{\d}}$ which in light-cone gauge
goes back to Plebanski's prepotential~$\Psi$~\cite{Ple},
\be
C_{\ad\bd\dt{\g}\dt{\d}}\ =\
\pa^+_{\ad}\pa^+_{\bd}\pa^+_{\dt{\g}}\pa^+_{\dt{\d}}\;\Psi \quad,
\ee
which is subject to the second-order equation~\gl{ple2}.

\medskip

In the complex notation, renaming $\Phi\to\phi$,
the $U(1)$ factor in \gl{break1} can be restored by
averaging over all cotangent structures.
In this manner, $\tilde{A}^{(3)}_2$ simplifies to
\bea
\int\!{d\th\over2\pi}\;\tilde{A}^{(3)}_2(k_1,k_2,k_3;\th)\ &=&\ -\sfrac14
\Bigl[\h_{A\bb}\,(k_1^A\,\bar{k}_2^\bb-\bar{k}_1^\bb\,k_2^A)\Bigr]^2 +\,
\sfrac12\,\e_{AB}\,k_1^A\,k_2^B\;\e_{\ba\bb}\,\bar{k}_1^\ba\,\bar{k}_2^\bb
\nonumber\\[1ex] \ &=&\ 
\sfrac32\;\e_{AB}\,\e_{\ba\bb}\;k_1^A\,k_2^B\,\bar{k}_1^\ba\,\bar{k}_2^\bb
\eea
which leads to a cubic interaction
\be
\cl_{\rm int}\ =\
\sfrac{\kappa}6\,\e^{AB}\,\e^{\ba\bb}\;
\phi\;\pa_A\pa_\ba\phi\;\pa_B\pa_\bb\phi \quad.
\ee
The corresponding equation of motion is
\be \label{ple1}
-\square\,\phi\ +\ \sfrac{\kappa}2\,\e^{AB}\,\e^{\ba\bb}\;
\pa_A\pa_\ba\phi\;\pa_B\pa_\bb\phi\ =\ 0
\ee
and is called Plebanski's {\it first\/} equation.
It also describes $2{+}2$ self-dual gravity but in a different
parametrization.
In particular, the metric of a self-dual spacetime is Ricci-flat and
K\"ahler, with a K\"ahler potential
\be \label{kaehler}
\Omega(y,\by)\ =\ \h_{A\ba}\,y^A \by^\ba\ +\ \phi(y,\by) \quad,
\ee
where $\phi$ must satisfy the second-order equation~\gl{ple1}.

\medskip

The Lagrangian may be generalized to curved Kleinian space~\cite{BGPPR}
by introducing
a K\"ahlerian metric background (plus a constant dilaton and axion).
The one-loop beta function equations then demand the K\"ahler metric
to be Ricci-flat, implying again a self-dual Riemann tensor~\cite{AFM}.
The deviation from flat $\R^{2,2}$ may therefore be parametrized
either by a nontrivial Weyl tensor expressed in~$\Psi$
or by the perturbation~$\phi$ of the flat K\"ahler potential,
in accordance with the string dynamics.
We learn that the arbitrariness of the Lorentz frame for the $N{=}2$ string
implies the equivalence of different descriptions of self-dual gravity,
in particular the equivalence of Plebanski's first and second equations
as a matter of gauge choice. For the remainder of this paper, we will
mainly use Plebanski's first equation, corresponding to complex
string variables, adapted to $SU(1,1)$ notation.
We conclude that, at tree-level, the $N{=}2$ closed string is indeed
identical to self-dual gravity.

\bigskip
\section{Self-dual gravity}

{\bf 3.1\ \  Self-dual gravity equations}
\smallskip

Let $M$ be an oriented four-manifold of class $\C^\o$ with a
nondegenerate metric $g$ of signature $(++-\,-)$ and a volume
element $\e$. Having the metric $g$, one can introduce the
Levi-Civita connection $\Gamma$, the Riemann tensor $\mr$ and
the Ricci tensor $Ric$. If $U$ is an open subset of $M$ and
$x^\mu : U\to\R^4$ are (local) coordinates on $U$, $\mu ,\nu
,...=1,...,4$, then these objects have components
$g=(g_{\mu\nu})$, $\Gamma =(\Gamma^\rho_{\mu\nu})$, $\mr
=(R^\rho_{\mu\s\nu})$, $Ric$~$= (R_{\mu\nu})=(R^\s_{\mu\s\nu})$,
and the volume element $\e$ has the form $$ \e
=\frac{1}{4!}\sqrt{\det g}\ \ve_{\mu_1...\mu_4} dx^{\mu_1}\wedge
...  \wedge dx^{\mu_4},  $$ where
$\ve_{\mu_1...\mu_4}$ is a skew-symmetric symbol with
$\ve_{1234}=1$.

\medskip

Let us consider the following equations on a metric $g=(g_{\mu\nu})$:
\be\label{(3.2)} *\mr =\mr \quad \Leftrightarrow\quad \frac{1}{2}
\e^{\mu_2\nu_2}_{\mu_1\nu_1}
R^\rho_{\s\mu_2\nu_2}=R^\rho_{\s\mu_1\nu_1},  \ee where $*$ is the
Hodge star operator and $\e^{\mu_2\nu_2}_{\mu_1\nu_1}:= \sqrt{\det
g}\ g^{\mu_2\rho}g^{\nu_2\s}\ve_{\rho\s\mu_1\nu_1}$. A metric $g$
satisfying eqs.\gl{(3.2)} is called a self-dual (SD) or
left-flat metric \cite{Pe, AHS, TW, Wa, HKLR}. It is easy to see that
SD metrics are Ricci-flat, i.e. they satisfy Einstein's vacuum field
equations.  Notice that the definition of self-duality will be
replaced by the definition of  anti-self-duality if we change the
orientation of the manifold $M$, e.g. from local coordinates $(x^1,
x^2, x^3, x^4)$ on $U$ we go over to $(x^1, x^2, x^3, - x^4)$.

\medskip

Equations \gl{(3.2)} can be rewritten in a simpler form of
equations on divergence-free vector fields. Namely, let $\e$ be a
volume form on an oriented 4-manifold $M$. Then for a vector field
$\varphi$ on $M$ a divergence of $\varphi$ is defined by
$$ \cl_\varphi\e = (\mbox{div}\varphi)\e , $$  where
$\cl_\varphi$ is the Lie derivative along $\varphi$. Thus,
``divergence-free" is a synonym for ``volume-preserving", and we
shall consider the algebra $\sdiff(M)$ of volume-preserving vector
fields on $M$. Now, let us take four pointwise linearly independent
vector fields $T_\a \in \sdiff(M)$ and suppose they satisfy the
self-duality equations \be\label{(3.4)}
\frac{1}{2}\ve^{\a_2\b_2}_{\a_1\b_1}[T_{\a_2}, T_{\b_2}]=[T_{\a_1},
T_{\b_1}], \ee where
$\ve^{\a_2\b_2}_{\a_1\b_1}:=\eta^{\a_2\a_3}\eta^{\b_2\b_3}
\ve_{\a_3\b_3\a_1\b_1}$, $\eta^{-1}=(\eta^{\a\b}) = $ diag$(+1, +1,
-1, -1)$ and $\a , \b,...$ are tangent (Lorentz) indices.  Let $f$ be
a scalar function defined by $f^2=4!\e (T_1, T_2, T_3, T_4)$, where
$\{T_\a\}$ is a solution of eqs.\gl{(3.4)}. Then one may define
a tetrad $e_\a$ and a (contravariant) metric $g^{-1}$ by formulae
\be\label{(3.5a)} e_\a :=f^{-1}T_\a\quad \Leftrightarrow\quad
e_\a^\mu =f^{-1}T_\a^\mu , \ee \be\label{(3.5b)}
g^{-1}:=f^{-2}\eta^{\a\b}T_\a T_\b \quad \Leftrightarrow\quad
g^{\mu\nu}=f^{-2}\eta^{\a\b}T_\a^\mu T_\b^\nu , \ee and the metric
\gl{(3.5b)} will be SD. Conversely, every SD metric arises in
this way, and eqs.\gl{(3.4)} are equivalent to
eqs.\gl{(3.2)}.  For detailed proofs and references see
e.g.~\cite{MW}.

\medskip

We call eqs.\gl{(3.4)} the {\it self-dual gravity} (SDG)
equations.  They are invariant under the transformations
\be\label{(3.6)}
T_\a\ \mapsto\ \d^0_\varphi T_\a = [\varphi , T_\a
], \ee where $\varphi$ is any divergence-free vector field on $M$,
i.e.  $\varphi\in\sdiff(M)$.  {}For discussion of Lorentz
transformations see~\cite{PBR}. Equations \gl{(3.4)} may be
interpreted as the self-dual Yang-Mills equations for Yang-Mills
fields with the gauge group SDiff$(M)$ of volume-preserving
transformations of $(M, \e )$ (see~\cite{MW} and references therein).

\bigskip
\bigskip\noindent
{\bf 3.2\ \ The first Plebanski equation}
\smallskip

Let us introduce complex divergence-free vector fields
\be\label{(3.7)} W_1:=\frac{1}{\sqrt{2}}(T_1-iT_2),\
W_2:=\frac{1}{\sqrt{2}}(T_3-iT_4),\ W_{\bar
1}:=\frac{1}{\sqrt{2}}(T_1+iT_2),\ W_{\bar
2}:=\frac{1}{\sqrt{2}}(T_3+iT_4), \ee i.e. $W_1, W_2, W_{\bar 1},
W_{\bar 2}\in\ $sdiff$^\C(M)=\ $sdiff$(M)\otimes\C$.  Then
eqs.\gl{(3.4)} may be rewritten in the form \be\label{(3.8a)}
[W_{\bar 1}, W_{\bar 2}]=0, \ee \be\label{(3.8b)} [W_{\bar 1},
W_{1}]-[W_{\bar 2}, W_{2}]=0, \ee \be\label{(3.8c)} [W_{1}, W_{2}]=0.
\ee We see that eqs.\gl{(3.8a)} and \gl{(3.8c)} have the
form of ``zero commutator" conditions. Notice that $\{W_A, W_{\bar
A}\}$ are null vector fields since $$ \eta_{1\bar 1}:=\eta (W_1,
W_{\bar 1})=1,\ \eta_{2\bar 2}:=\eta (W_2, W_{\bar 2})=-1, $$ $$ \eta
(W_A, W_B)=\eta (W_{\bar A}, W_{\bar B})= \eta(W_1, W_{\bar 2})=
\eta(W_2, W_{\bar 1})=0, $$ where $A,B,...=1,2,\ \bar A,\bar B,
...=1,2$.

\medskip

We have defined complex vector fields \gl{(3.7)} so that
$W_{\bar A}=\overline{W_A}$. Notice that eqs.\gl{(3.8a)} and
\gl{(3.8c)} are invariant w.r.t. transformations from the
complexification SDiff$^\C(M)$ of the group SDiff$(M)$. We shall
consider transformations from SDiff$^\C(M)$ such that after their
action we shall have $W_{\bar A}\ne\overline{W_A}$  but the metric
will be real.  These complex transformations may be used for the
partial fixing of a coordinate system. Namely, one can always
introduce complex coordinates
$y^A$, $y^{\bar A}:=\overline{(y^A)}={\bar y}^{\bar A}$
on $U\subset M$ so that $W_{\bar 1}$ and $W_{\bar 2}$ become
coordinate derivatives (Frobenius theorem), i.e.  \be\label{(3.9a)}
W_{\bar A}=\p_{\bar A}:=\frac{\p}{\p y^{\bar A}}, \ee and
eq.\gl{(3.8a)} is identically satisfied. Then one can solve
eq.\gl{(3.8b)} by choosing a gauge \be\label{(3.9b)}
W_1=\p_{1}\p_{\bar 2}\Omega\p_2 -  \p_{2}\p_{\bar 2}\Omega\p_1\
,\quad W_2=\p_{1}\p_{\bar 1}\Omega\p_2 - \p_{2}\p_{\bar 1}\Omega\p_1\
, \ee where $\Omega (y^A, y^{\bar A})$ is a real-valued scalar
function, $\p_A:={\p}/{\p y^{A}}$. Substituting \gl{(3.9b)}
into eq.\gl{(3.8c)}, we obtain the first Plebanski equation
\be\label{(3.10)} \p_{1}\p_{\bar 2}\Omega\p_{2}\p_{\bar 1}\Omega -
\p_{1}\p_{\bar 1}\Omega \p_{2}\p_{\bar 2}\Omega =1.  \ee Thus, the
SDG equations \gl{(3.8a)}-\gl{(3.8c)} can be reduced to one equation
\gl{(3.10)} on a scalar function $\Omega$ which coincides with eq.\gl{ple1}
with $\kappa{=}1$ after shifting $\Omega$ by formula~\gl{kaehler}.

\medskip

 We introduce the antisymmetric $\ve$-symbols, $$
\ve_{12}=-\ve_{21}=\ve_{\bar 1\bar 2}=-\ve_{\bar 2\bar 1}=1,\quad
\ve^{12}=-\ve^{21}=\ve^{\bar 1\bar 2}=-\ve^{\bar 2\bar 1}=-1, $$
\be\label{(3.11)} \ve_A^{\bar B}=\eta^{\bar BC}\ve_{CA},\
\ve^B_{\bar A}=\eta^{B\bar C}\ve_{\bar CA}\quad \Rightarrow\quad
\ve^2_{\bar 1}=\ve^1_{\bar 2}=\ve^{\bar 2}_1=\ve^{\bar 1}_2=1. \ee
By using \gl{(3.11)}, eqs. \gl{(3.9a)} -- \gl{(3.10)} can be rewritten
in the form \be\label{(3.12a)}
W_{\bar A}=\p_{\bar A}, \quad W_A=\ve^{\bar
B}_{A}\ve^{CB}\p_B\p_{\bar B}\Omega\p_C , \ee \be\label{(3.12b)}
\ve^{AB}\ve^{\bar A\bar B}\p_{A}\p_{\bar A}\Omega\p_{B}\p_{\bar
B}\Omega =-2.  \ee

\medskip

The vector fields \gl{(3.12a)} are divergence-free with respect
to the volume form $\epsilon =dy^1\wedge dy^2\wedge dy^{\bar 1}\wedge
dy^{\bar 2}$ and coincide with a tetrad because $f=1$.  Dual basis of
$\{W_A,W_{\bar A}\}$ has the form $$ \th^1=\p_{\bar
1}\p_A\Omega dy^A,\ \th^2=-\p_{\bar 2}\p_A\Omega dy^A,\ \th^{\bar
A}=dy^{\bar A}. $$ The above choice of coordinates, a tetrad and a
dual tetrad is convenient because the metric $g$ in this gauge takes
the K\"ahler form \be\label{(3.14)} g=\eta_{A\bar
B}\th^A\th^{\bar B}=2\p_A\p_{\bar B}\Omega dy^A dy^{\bar B} \ee with
the K\"ahler potential $\Omega$.

\bigskip
\bigskip\noindent
{\bf 3.3\ \ Linear systems for the SDG equations}
\smallskip

In Sect.3.2 it has been shown that the first Plebanski equation
\gl{(3.10)} can be obtained from the ASDG equations
\gl{(3.8a)}-\gl{(3.8c)} by fixing a gauge. That is why we shall consider
`Lax pairs' for equations \gl{(3.8a)}-\gl{(3.8c)}, and Lax pairs for
eq.\gl{(3.10)} will be obtained by substituting the tetrad
\gl{(3.12a)}. For the sake of simplicity we shall consider real
analytic solutions of eqs.\gl{(3.8a)}-\gl{(3.8c)} and \gl{(3.10)}.

\medskip

Equations \gl{(3.8a)}-\gl{(3.8c)} can be obtained as the compatibility
conditions of the following linear system of equations:
$$ \cl^+_1\phi_+^a := (W_{\bar 1} - \l W_2)\phi_+^a
=0, $$ $$ \cl^+_2\phi_+^a := (W_{\bar 2} - \l
W_1)\phi_+^a =0, $$ \be\label{(3.15)} \cl^+_3\phi_+^a :=
\p_{\bar\l}\phi_+^a =0, \ee where $\l$ is the complex `spectral
parameter', and $\phi^a_+(x ,\l )$ are smooth functions for $x\in U$,
$|\l |\le 1$ and holomorphic in $\l$ for $|\l |<1$, $a,b,...= 1,2,3$
.  Indeed, the compatibility conditions \be\label{(3.16)} [\cl^+_a,
\cl^+_b]=0 \ee of eqs.\gl{(3.15)} are identical to the SDG
equations \gl{(3.8a)}-\gl{(3.8c)}.

\medskip

{}For the same equations \gl{(3.8a)}-\gl{(3.8c)} one can write another
linear system $$ \cl^-_1\phi_-^a := ({\l}^{-1}W_{\bar
1} - W_2)\phi_-^a =0, $$ $$ \cl^-_2\phi_-^a :=
({\l}^{-1}W_{\bar 2} - W_1)\phi_-^a =0, $$ \be\label{(3.17)}
\cl^-_3\phi_-^a := \p_{\bar\l}\phi_-^a =0, \ee where $|\l |>0$ and
functions $\phi^a_-(x,\l )$ are smooth in $x\in U$, $|\l |\ge 1$ and
holomorphic in $\l$ for $|\l |>1$. The compatibility conditions
\be\label{(3.18)} [\cl^-_a, \cl_b^-]=0 \ee of eqs.\gl{(3.17)}
are identical to the SDG equations \gl{(3.8a)}-\gl{(3.8c)}.

\medskip

The auxiliary `spectral parameter' $\l$ in the linear systems
\gl{(3.15)} and \gl{(3.17)} is connected with the group
$SU(1,1)$. This group acts on the sphere $S^2\simeq \C P^1$, and the
Riemann sphere decomposes into the disjoint union \be\label{a)}
S^2=H^2_+\cup S^1\cup H^2_-  \ee of orbits of $SU(1,1)$, where
$$H^2:=H^2_+\cup H^2_-  $$ is the two-sheeted
hyperboloid, $H^2_+\simeq SU(1,1)/U(1)$ is the upper half of the
hyperboloid and $H^2_-\simeq SU(1,1)/U(1)$ is the lower half of the
hyperboloid. Using the stereographic projection, one can identify
$H^2_+$ with the open disk $|\l |<1$ in $\C$ and $H^2_-$ with the
domain $|\l |>1$ in $\C\cup\infty$.

\medskip

One may consider the closures $$ \bar H^2_+=H^2_+\cup
S^1=\{\l\in\C :  |\l |\le 1\},\ \bar H^2_-=H^2_-\cup
S^1=\{\l\in\C\cup\infty :  |\l |\ge 1\}.   $$ Then $\l\in \bar
H^2_+$ in \gl{(3.15)},  $\l\in \bar H^2_-$ in \gl{(3.17)}
and $S^1$ is a common boundary of the spaces $\bar H^2_+$ and $\bar
H^2_-$.  Notice that the metric \be ds^2=\frac{4d\l
d\bar\l}{(1-\l\bar\l)^2} \ee on $H^2$ is invariant under the action
of the group $SU(1,1)$ and singular at $|\l |=1$. This is a metric of
the Poincar\'e model of Lobachevskii geometry.

\vfill\eject
\bigskip\noindent
{\bf 3.4\ \ Twistors, complex structures and the SDG equations}
\smallskip

Let us consider a bundle $\cz\to M$ of complex structures over a
4-manifold $M$. Following ref.~\cite{AHS}, we shall call $\cz$ a
twistor space of $M$. Since we are interested in local geometry of
the manifold $M$, we shall consider the restriction of the twistor
bundle $\cz\to M$ to an open subset $U\subset M$ and put $\cp
:=\cz|_U$.

\medskip

Recall that on any oriented SD manifold $M$ there exists a family of
complex structures covariantly constant w.r.t. the Levi-Civita
connection, and they are parametrized by the two-sheeted hyperboloid
$H^2$ described in Sect.3.3, where $$ H^2_+\simeq
SO(2,2)/U(1,1)\simeq SU(1,1)/U(1)\simeq SL(2,\R)/SO(2), $$ $$
H^2_-\simeq SO(2,2)/U(1,1)\simeq SU(1,1)/U(1)\simeq SL(2,\R)/SO(2) $$
are two orbits of the group $SO(2,2)$ acting on complex structures
$J\in SO(2,2)$.  Therefore, in our case the twistor space $\cp$ of
$U$ coincides, as a smooth manifold, with the direct product
$$\cp \simeq U\times H^2 =(U\times H^2_+)\cup
(U\times H^2_-)=\cp_+\cup\cp_-, $$ i.e. $\cp$ is the disjoin union
of $\cp_+=U\times H^2_+$ and $\cp_-=U\times H^2_-$.

\medskip

The vector fields $\cl^+_a$ from the linear system \gl{(3.15)}
span the (0,1) tangent space of $\cp_+$, and the vector fields
$\cl^-_a$ from \gl{(3.17)} span the (0,1) tangent space of
$\cp_-$.  The compatibility conditions of these linear systems are
the integrability conditions of an almost complex structure on the
twistor space $\cp$. {}From the other hand, these conditions are
equivalent to the condition of self-duality of the metric $g$ on
$U\subset M$, which is the reformulation to the ultrahyperbolic case
of the well-known twistor correspondence between  SD geometry of $U$
and complex geometry of $\cp$~\cite{Pe, AHS}.  Notice that if an
almost complex structure on $\cp$ is integrable, then $\phi^a_\pm$
may be taken as complex analytic coordinates on $\cp_\pm\subset \cp$.
Moreover, from \gl{(3.15)} and \gl{(3.17)} it is clear
that one may always take $\phi^3_+(x,\l )=\l ,\quad  \phi^3_-(x,\l
)=\l^{-1}$.

\medskip

The space $$ \cp_0=U\times S^1  $$ is the $S^1$
bundle over $U$ of real anti-self-dual null bivectors
$\cl^0_1\wedge\cl^0_2$, modulo scale,  where \be\label{(3.22)}
\cl_1^0=W_{\bar 1}-\l W_2,\quad \cl_2^0=\l^{-1}W_{\bar 2}-W_1,  \ee
$\l\in S^1$.  Put another way, $\cp_0$ is a set of pairs $(x,
\cl^0_1(x) \wedge\cl^0_2(x))$, where $x\in U$, and $$
\cl^0_1\wedge\cl^0_2=\l^{-1}W_{\bar 1}\wedge W_{\bar 2} + (W_1\wedge
W_{\bar 1}-W_2\wedge W_{\bar 2})-\l W_1\wedge W_{2} $$ is an
 anti-self-dual null bivector at this point, parametrized by $\l\in
S^1$. If we introduce  the spaces $$
\bar\cp_+:=U\times \bar H^2_+, \quad \bar\cp_-:=U\times \bar H^2_-,
$$ then $\cp_0=\bar\cp_+\cap\bar\cp_-$ and our linear systems
\gl{(3.15)}, \gl{(3.17)} are defined on the subsets
$\bar\cp_+ $, $\bar\cp_-$ of the space $$ \tilde\cp
:= \cp_+\cup\cp_0\cup\cp_-\simeq U\times S^2.   $$

\medskip

The space $\cp_0=\bar\cp_+\cap\bar\cp_-$ is the common domain of the
coordinates $\phi^a_+$ and $\phi^a_-$. Therefore there exist smooth
functions $f_{+-}^a$ such that on $\cp_0$ we have \be\label{(3.25)}
\phi^a_+= f_{+-}^a(\phi^b_-) .  \ee On the space $\tilde\cp$ one can
introduce a map $\tau :  \tilde\cp\to\tilde\cp$ called a {\it real
structure}. It is an antiholomorphic involution, defined by the
formula
\be\label{(3.26a)} \tau (\phi^1(x,\l ), \phi^2(x,\l ),
\phi^3(x,\l ))= (\phi^1(\bar x,\bar\l^{-1} ), \phi^2(\bar
x,\bar\l^{-1} ), \phi^3(\bar x,\bar\l^{-1})),
\ee
where $\phi^a(x,\l )$ are local coordinates on $\tilde\cp$,
$\bar x=x$.  This involution
takes the complex structure on $\tilde\cp$ to its conjugate.  We may
choose the coordinates $\phi^a_\pm$ on $\bar\cp_{\pm}$ in such a way
that \be\label{(3.26b)} \phi^1_-=\tau (\overline{\phi^2_+}),\
\phi^2_-=\tau (\overline{\phi^1_+}),\ \phi^3_-=\tau
(\overline{\phi^3_+}).  \ee Considering these conditions on $\cp_0$
and substituting them into {\gl{(3.25)}}, we obtain the reality
conditions on functions $\phi^a_{\pm}$.

\medskip

There are fixed points of the action of $\tau$ on $\tilde\cp$, and
they form a three-dimensional real manifold called a real twistor
space $\ct (U)$ of $U$. In the flat case this manifold coincides with
the projective space $\R P^3$~\cite{Wo, BGPPR}.  The space $\cp_0$ is
fibred over $\ct (U)$ by real 2-manifolds called $\b$-surfaces, and
the vector fields \gl{(3.22)} span their tangent spaces.
Coordinates on the real twistor space $\ct (U)$ may be introduced in
the following way. Consider three functions $\phi^a_0(x,\l )$ on
$\cp_0$ which are real analytic for $x\in U$, $\l =\exp (i\gamma )\in
S^1$. So they extend holomorphically in $\l$ to a neighbourhood
$\U_0$ of $U\times S^1$ in $\tilde\cp = U\times S^2$. Then consider
linear differential equations \be\label{(3.27a)}
\cl^0_1\phi^a_0=(W_{\bar 1}-\l W_2)\phi^a_0=0,  \ee
\be\label{(3.27b)}
\cl^0_2\phi^a_0=(\l^{-1}W_{\bar 2}-W_1)\phi^a_0=0,
\ee where $\cl^0_A$ are vector fields on $\cp_0$ defined by formulae
\gl{(3.22)}.  The compatibility conditions of the linear system
\gl{(3.27a)}, \gl{(3.27b)} are identical to the SDG equations.  Therefore
eqs.\gl{(3.27a)}, \gl{(3.27b)} always have solutions if $\{W_A, W_{\bar A}\}$
satisfy eqs.\gl{(3.8a)}-\gl{(3.8c)}.  Moreover, one may always take
$\phi^3_0=\l$ and impose on $\phi^1_0, \phi^2_0$ the reality
condition such that \be\label{(3.28)}\overline{\phi^1_0(x,\l
)}=\phi^2_0(x,\l ). \ee Such functions $(\phi^1_0, \phi^2_0,
\phi^3_0)=(\phi^1_0, \overline{\phi^1_0}, \exp(i\gamma ))$ may be
considered as coordinates on real twistor space $\ct (U)$ of $U$.
Notice that extensions of these real analytic functions $\phi^a_0$ to
a neighbourhood $\U_0$ of $U\times S^1$  are holomorphic coordinates
on $\U_0$ satisfying the reality conditions
$$\overline{\phi^1_0 (x, \bar\l^{-1})}=\phi^2_0(x,\l
),$$ which are reduced to the condition \gl{(3.28)} on
$\cp_0$.

\medskip

To sum up, by virtue of the twistor correspondence  all the
information about a SD metric on $U$ is encoded in a complex
structure of the twistor space $\cp$ of $U$, which is defined by the
linear systems \gl{(3.15)}, \gl{(3.17)} and complex
coordinates $\phi^a_\pm$. {}From the other hand, according to the
Kodaira-Spencer deformation theory, all the information about the
complex structure of $\cp$ can be extracted from the transition
functions $f^a_{+-}$ on $\cp_{0}$.  But for preserving readability we
shall not develop this correspondence further.

\vfill\eject
\bigskip\noindent
{\bf 3.5\ \ The second Plebanski equation}
\smallskip

To describe the second Plebanski equation, it is necessary to go to a
real null tetrad $\{X_A, X_{\tilde A}\}$ on $U\subset M,\
A,B,...=1,2, \tilde A, \tilde B,...=1,2$. For this we set
\be\label{(3.30)} \l =\frac{i-\zeta}{i+\zeta} \ee in
eqs.\gl{(3.15)} and \gl{(3.17)}. The transformation
\gl{(3.30)} is a linear-fractional transformation of the
complex plane that carries the unit disk $|\l |<1$ to the upper
half-plane Im$\zeta >0$, the domain $|\l |>1$ to the lower half-plane
Im$\zeta <0$ and the circle $|\l |=1$ to the real axis Im$\zeta =0$.
The metric on $H^2$ in the coordinates $\zeta , \bar\zeta$ takes the
form \be ds^2=-\frac{4d\zeta d\bar\zeta}{(\zeta -\bar\zeta )^2}.  \ee
This is a metric of Klein's model of Lobachevskii geometry.  It is
invariant under the action  of the group $SL(2,\R)$ and singular when
Im$\zeta =0$ (real axis).

\medskip

It is easy to show that after the above transformation $\l\mapsto
\zeta$, eqs.\gl{(3.15)} may be rewritten as \be\label{(3.31)}
\frac{1}{(i+\zeta )}(X_1-\zeta X_{\tilde 2})\chi^a_+=0,\
\frac{1}{(i+\zeta )}(X_2-\zeta X_{\tilde 1})\chi^a_+=0,\
\p_{\bar\zeta}\chi^a_+=0, \ee where Im$\zeta\ge 0$ and
$$ X_1=\frac{1}{\sqrt{2}}(T_2+T_4),\
X_2=\frac{1}{\sqrt{2}}(T_3-T_1),\ X_{\tilde
1}=\frac{1}{\sqrt{2}}(T_2-T_4),\ X_{\tilde
2}=\frac{1}{\sqrt{2}}(T_3+T_1) $$ are null vector fields since \be
\eta (X_1,X_{\tilde 1} ) = -\eta (X_2,X_{\tilde 2} )=1,\ \eta
(X_A,X_B)=\eta (X_{\tilde A},X_{\tilde B})= \eta (X_1,X_{\tilde
2})=\eta (X_2,X_{\tilde 1})=0.  \ee Analogously,
eqs.\gl{(3.17)} may be rewritten in the form \be\label{(3.33)}
\frac{1}{(i-\zeta )}(X_1-\zeta X_{\tilde 2})\chi^a_-=0,\
\frac{1}{(i-\zeta )}(X_2-\zeta X_{\tilde 1})\chi^a_-=0,\
\p_{\bar\zeta}\chi^a_-=0, \ee where Im$\zeta\le 0$. Functions
$\chi^a_+(x,\zeta )$ and $\chi^a_-(x,\zeta )$ are real analytic for
$x\in U$, Im$\zeta\ge 0$ and $x\in U$, Im$\zeta\le 0$, respectively.
The functions $\chi^a_+$  are holomorphic in $\zeta$ for Im$\zeta >0$
and $\chi^a_-$  are holomorphic in $\zeta$ for Im$\zeta <0$.
Moreover, one may always take $\chi^3_+=\chi^3_-=\zeta$.  Notice that
the reality condition \gl{(3.26b)} takes the form
$$ (\overline{\chi^1_+(x,\bar\zeta
)},\overline{\chi^2_+(x,\bar\zeta )}, \overline{\chi^3_+(x,\bar\zeta
)})= (\chi^2_-(x,\zeta ), \chi^1_-(x,\zeta ), \chi^3_-(x,\zeta )).
$$

\medskip

The compatibility  conditions of each of the linear systems
\gl{(3.31)} and \gl{(3.33)} are reduced to the equations
\be\label{(3.35a)} [X_1, X_2]=0, \ee \be\label{(3.35b)} [X_{\tilde 1},
X_1]-[X_{\tilde 2}, X_2]=0, \ee \be\label{(3.35c)} [X_{\tilde 1},
X_{\tilde 2}]=0, \ee which are the SDG equations \gl{(3.4)}
rewritten in terms of real divergence-free vector fields $\{X_A,
X_{\tilde A}\}$. Again, as in the case of complex divergence-free
vector fields, two equations from eqs.\gl{(3.35a)}-\gl{(3.35c)} has the form
of ``zero commutator" conditions and one can choose {\it real}
coordinates such that $\{X_A\}$ or $\{X_{\tilde A}\}$ coincide with
coordinate derivatives. Then for two remaining vector fields one can
choose a parametrization such that eq.\gl{(3.35b)} will be
identically satisfied, and the SDG equations will be reduced to one
equation on a scalar function.

\medskip

Let us choose real coordinates $(z^A, z^{\tilde A})$ so that
\be\label{(3.36a)} X_A=\p_A:=\frac{\p}{\p z^A}, \ee and take vector
fields $\{X_{\tilde A}\}$ in the form \be\label{(3.36b)} X_{\tilde
1}=\p_2\p_{\tilde 1}\Omega\p_{\tilde 2} - \p_2\p_{\tilde
2}\Omega\p_{\tilde 1},\quad X_{\tilde 2}=\p_1\p_{\tilde
1}\Omega\p_{\tilde 2} - \p_1\p_{\tilde 2}\Omega\p_{\tilde 1}, \ee
where $\Omega =\Omega (z^A, z^{\tilde A})$ is a real-valued scalar
function and $\p_{\tilde A}:=\p /\p z^{\tilde A}$. Then
eq.\gl{(3.35b)} is identically satisfied, and
eq.\gl{(3.35c)} is reduced to the first Plebanski equation
$$ \p_1\p_{\tilde 2}\Omega\p_2\p_{\tilde 1}\Omega -
\p_1\p_{\tilde 1}\Omega\p_2\p_{\tilde 2}\Omega =1 $$ in the real
null coordinates $(z^A, z^{\tilde A})$.

\medskip

Using the same coordinates $(z^A, z^{\tilde A})$, instead of
\gl{(3.36a)}, \gl{(3.36b)} one can choose the following tetrad:
\be\label{(3.38)} X_A=\p_A ,\ X_{\tilde 1}=\p_{\tilde
1}+\p_1\p_2\Psi\p_2 -\p_2^2\Psi\p_1 ,\ X_{\tilde 2}=\p_{\tilde
2}+\p_1^2\Psi\p_2-\p_1\p_2\Psi\p_1\ .  \ee Then
eqs.\gl{(3.35a)}, \gl{(3.35b)} are identically satisfied,
and eq.\gl{(3.35c)} is reduced to the second Plebanski equation
\be\label{(3.39)} \p_1\p_{\tilde 1}\Psi -\p_2\p_{\tilde 2}\Psi
+\p_1\p_2\Psi\p_1\p_2\Psi - \p^2_1\Psi\p^2_2\Psi =0 \ee on a
real-valued scalar function $\Psi (z^A, z^{\tilde A})$.
This equation coincides with eq.\gl{ple2} after introducing the
coupling constant $\kappa$ by changing $\Psi\to\kappa\Psi$.

\medskip

The vector fields \gl{(3.36a)},  \gl{(3.36b)}, \gl{(3.38)} are
divergence-free with respect to the volume form $\e =dz^1\wedge
dz^2\wedge dz^{\tilde 1}\wedge dz^{\tilde 2}$ and dual basis of the
tetrad \gl{(3.38)} has the form $$
\sigma^1=dz^1+\p_1\p_2\Psi dz^{\tilde 2}+\p^2_2\Psi dz^{\tilde 1}, \
\sigma^2=dz^2-\p_1\p_2\Psi dz^{\tilde 1}-\p^2_1\Psi dz^{\tilde 2}, \
\sigma^{\tilde A}=dz^{\tilde A}.  $$ In terms of $\Psi$ the metric
 $g$ has the form
\be\label{(3.40b)} g=\eta_{A\tilde
B}\sigma^{A}\sigma^{\tilde B} = 2(dz^1dz^{\tilde 1} - dz^2dz^{\tilde
2} + \p^2_2\Psi dz^{\tilde 1} dz^{\tilde 1}+2\p_1\p_2\Psi dz^{\tilde
1} dz^{\tilde 2} + \p^2_1\Psi dz^{\tilde 2} dz^{\tilde 2}). \ee

\medskip

{\bf Remark.} The Plebanski equations are special cases of the SDG
equations appeared after the gauge fixing and solving two equations
from eqs.\gl{(3.4)}. The first Plebanski equation is an
analogue of the SD Yang-Mills equations in the so-called Yang gauge,
and the second Plebanski equation is an analogue of the SD Yang-Mills
equations in the so-called Leznov gauge.

\vfill\eject

\section{Symmetries of self-dual gravity}

{\bf 4.1\ \ Symmetries of the SDG equations from the twistor viewpoint}
\smallskip

We consider the SDG equations \gl{(3.8a)}-\gl{(3.8c)}.
The linearization of these
equations has the form
$$
[\d W_{\bar 1}, W_{\bar 2}]+ [W_{\bar 1}, \d W_{\bar 2}]=0, \quad
[\d W_{1}, W_{2}]+ [W_{1}, \d W_{2}]=0, $$
\be\label{(4.1)}
[\d W_{\bar 1}, W_{1}]+ [W_{\bar 1}, \d W_{1}]-[\d W_{\bar 2}, W_{2}]
-[ W_{\bar 2},\d W_{2}]=0.
\ee
To find (infinitesimal) {\it symmetries} of eqs.\gl{(3.8a)}-\gl{(3.8c)}
means to find solutions $\d W_A$, $\d W_{\bar A}$ of eqs.\gl{(4.1)} for all
given solutions $W_A$, $W_{\bar A}$ of eqs.\gl{(3.8a)}-\gl{(3.8c)}.

\medskip

As discussed in Sect.3.3, eqs.\gl{(3.8a)}-\gl{(3.8c)} can be rewritten
as eqs.\gl{(3.16)} or
eqs.\gl{(3.18)} depending on $\l\in\bar H^2_+$ or $\l\in\bar H^2_-$. Since
the vector fields $\cl^\pm_3$ trivially commute with the vector
fields  $\cl^\pm_A$, one may consider only the following
equations:  \be\label{(4.2a)} \bar\cp_+:\ \quad  [\cl^+_1, \cl^+_2]=0,
\ee \be\label{(4.2b)}\bar\cp_-:\ \quad  [\cl^-_1, \cl^-_2]=0,
\ee defined on the subsets $\bar\cp_+=U\times\bar H^2_+$
and $\bar\cp_-= U\times\bar H^2_-$ of the space $\tilde \cp$ (see
Sect.3.4). Accordingly, equations \gl{(4.1)} can be rewritten in the
form of the following equations on $\bar\cp_\pm$:
\be\label{(4.3a)}
 [\d\cl^+_1, \cl^+_2]+ [\cl^+_1,\d \cl^+_2]=0, \ee
\be\label{(4.3b)}
[\d\cl^-_1, \cl^-_2]+[\cl^-_1,\d \cl^-_2]=0, \ee where
\be\label{(4.3c)}\d\cl^+_1:=\d W_{\bar 1}-\l\d W_2, \quad \d\cl^+_2:= \d
W_{\bar 2}-\l\d W_1, \ee
\be\label{(4.3d)}\d\cl^-_1:=\frac{1}{\l}\d
W_{\bar 1}-\d W_2, \quad \d\cl^-_2:=\frac{1}{\l}\d W_{\bar 2}-\d
W_1.  \ee
Solutions of eqs.\gl{(4.3a)} on $\bar\cp_+$ are related to solutions of
eqs.\gl{(4.3b)} on $\bar\cp_-$ by
\be\label{(4.4)}
\d\cl^+_A=\l\d\cl^-_A \ee
on the overlap $\cp_0=\bar\cp_+\cap\bar\cp_-$,
since eqs.\gl{(4.3a)} and \gl{(4.3b)} on the subsets of $\tilde\cp$ encode
the same equations \gl{(4.1)} on $U\subset M$.

\medskip

General solutions of eqs.\gl{(4.3a)} and \gl{(4.3b)} have the form
$$
\d_\psi\cl^+_A=[\psi_+, \cl^+_A],
$$
\be\label{(4.5)}
\d_\psi\cl^-_A=[\psi_-, \cl^-_A],
\ee
where $\psi_+=\psi_+^\mu (x,\l )\p_\mu$ is a vector field on
$U\times\bar H^2_+$, the
components of which are smooth for $x\in U$, $\l\in\bar H^2_+$ and
holomorphic in $\l$ for $|\l |<1$, and $\psi_-= \psi_-^\mu (x,\l
)\p_\mu$ is a smooth vector field on $U\times\bar H^2_-$, the
components of which are holomorphic in $\l$ for $|\l |>1$. These
vector fields should preserve a volume form $\epsilon$ on $U$.

\medskip

By substituting the solutions \gl{(4.5)} into eqs.\gl{(4.4)},
we obtain the equations
$$
[\psi_+-\psi_-, \cl^+_A]=0
$$
on the intersection $\cp_0$ of $\bar\cp_+$ and $\bar\cp_-$.
These equations mean that the vector fields $\psi_+$ and $\psi_-$ are
not completely arbitrary.

\medskip

Symmetries of eqs.\gl{(4.2a)}, \gl{(4.2b)} and therefore of the SDG equations
\gl{(3.8a)}-\gl{(3.8c)} may
be described as follows. Consider the space
$\tilde\cp\simeq U\times\C P^1$, the subsets
$\bar\cp_\pm\subset\tilde\cp$ and the nonempty intersection
$\cp_0=\bar\cp_+\cap \bar\cp_-$ (see Sect.3.4).  On $\cp_0=U\times
S^1$ we introduce a vector field $\psi =\psi^\mu\p_\mu$
satisfying the equations \be\label{(4.7)} [\psi , \cl^+_A]=0 \ee and
preserving a volume form $\epsilon$ on $U$. Then by expanding
$\psi$ in a Fourier series in $\l =\exp (i\gamma )\in S^1$ we
have
\be\label{(4.8a)}\psi = \sum^\infty_{n=-\infty}\l^n\psi^n,
\ee
\be\label{(4.8b)}\psi =\psi_+-\psi_-,\ee
\be\label{(4.8c)}\psi_+:=\psi_+^0 + \sum^\infty_{n=1}\l^n\psi^n, \quad
\psi_-=\sum^\infty_{n=0}\l^{-n}\psi_-^n:=\psi_-^0 -
\sum^\infty_{n=1}\l^{-n}\psi^{-n}, \quad
\psi^0_+-\psi^0_-=\psi^0,\ee
where $\psi^n=\psi^{n,\mu}\p_{\mu}$ are vector fields on $U$. As a
function of $\l$, each component $\psi_+^\mu$ of $\psi_+$ is the limit
of a holomorphic function on the open disk $|\l |<1$ and $\psi_-^\mu$
is the limit of a holomorphic function on the exterior $|\l |>1$,
including the point $\l =\infty$. Put another way, $\psi_+$
extends continuously to a vector field on $\bar \cp_+$ with
components holomorphic in $\l\in H^2_+$, and $\psi_-$ extends
continuously to a vector field on $\bar \cp_-$ with components
holomorphic in $\l\in H^2_-$.  The splitting \gl{(4.8a)}-\gl{(4.8c)}
of $\psi$ is
unique up to
\be\label{(4.9)}
\psi^0_+\mapsto\psi^0_++\varphi ,\quad
\psi^0_-\mapsto\psi^0_-+\varphi ,\ee
for some vector field $\vp =\vp^\mu (x)\p_\mu \in\ $sdiff$(M)$.
Using the transformations \gl{(4.9)}, one can always choose a gauge in
which $\psi^0_+$ or $\psi^0_-$ is zero. Finally, using
$\psi_\pm$ from \gl{(4.8a)}-\gl{(4.8c)}
we define the transformations \gl{(4.5)} of the
vector fields $\cl^\pm_A$. By construction, these $\d_\psi\cl^\pm_A$
satisfy eqs.\gl{(4.3a)}, \gl{(4.3b)} and \gl{(4.4)}.

\medskip

It is not difficult to verify that if vector fields $\psi$ and
$\eta$ on $\cp_0$ satisfy eqs.\gl{(4.7)},  then the vector field
$[\psi ,\eta ]$ will also satisfy eqs.\gl{(4.7)} by virtue of the
Jacobi identities.  Therefore, the space of vector fields
satisfying eqs.\gl{(4.7)} forms an algebra with the standard commutator
of vector fields.  These vector fields can be considered as
representatives of `free' vector fields on the real twistor space
$\ct (U)$ of $U$. Notice that to impose reality conditions
on symmetries of equations \gl{(3.8a)}-\gl{(3.8c)} and \gl{(4.2a)},
\gl{(4.2b)} it is necessary to know
reality conditions for vectors from a tetrad $\{W_A,
W_{\bar A}\}$. For example, if $W_{\bar A}=
\overline{W_A}$, then on vector fields $\psi_{\pm}$ from formulae
\gl{(4.8a)}-\gl{(4.8c)} one can impose the following reality conditions:
$$
\psi_+(x,\l )=\overline{\psi_-(x, \bar\l^{-1})} \quad
\Longleftrightarrow\quad \psi^n_+=\overline{\psi^n_-},
$$
where $n=0,1,...\ $. Then we shall have $\d_\psi W_{\bar A}=
\overline{\d_\psi W_{A}}$.

\medskip

Suppose that not only the vector field $\psi$, but also the vector
fields
$\psi_\pm$ from \gl{(4.8a)}-\gl{(4.8c)} satisfy eqs.\gl{(4.7)}, i.e.
\be\label{(4.11)} [\psi_\pm ,\cl^+_A]=0.
\ee

Then from formulae \gl{(4.3c)}, \gl{(4.3d)} and \gl{(4.5)} we see that for such
$\psi_\pm$ the symmetry transformations are trivial, $\d_\psi
W_A= \d_\psi W_{\bar A}=0$. Vector fields $\psi_+$ and
$\psi_-$ satisfying eqs.\gl{(4.11)} are in a one-to-one correspondence
with holomorphic vector fields on $\cp_+$ and $\cp_-$,
respectively. Indeed, $\cl^+_a$ are vector fields of type (0,1)
on $\cp_+$ (see Sect.3.4) and therefore
have the form
$$
\cl^+_a = \cl^{+\bar b}_a \frac{\p}{\p\phi^{\bar b}_+}.
$$
{}From eqs.\gl{(4.11)} for components of a vector field
$$
\psi_+=\psi_+^a\frac{\p}{\p\phi^{a}_+}+\psi_+^{\bar a}
\frac{\p}{\p\phi^{\bar a}_+}
$$
we obtain
\be\label{(4.12c)}
\frac{\p}{\p\phi^{\bar b}_+}\psi^a_+=0,
\ee
\be\label{(4.12d)}
\psi^{\bar b}_+\frac{\p}{\p\phi^{\bar b}_+}\cl^{+\bar a}_c -
\cl^{+\bar b}_c\frac{\p}{\p\phi^{\bar b}_+}\psi_+^{\bar a} +
\psi^b_+\frac{\p}{\p\phi^{b}_+} \cl^{+\bar a}_c =0.
\ee
Equations \gl{(4.12c)} mean that the vector field $\psi_+$ has
arbitrary holomorphic components $\psi^a_+$  in a
complex holonomic basis $\{\p /\p\phi^a_+, \p /\p\phi^{\bar
a}_+\}$ of vector fields on $\cp_+$ and fixed (0,1)
components $\psi^{\bar a}_+$ determined by eqs.\gl{(4.12d)}.
Analogously, solutions $\psi_-$ of eqs.\gl{(4.11)} are in a
one-to-one correspondence with holomorphic vector fields on
$\cp_-$.

\medskip

{\bf Remark}. The quotient space of the space of solutions $\psi$
to eqs.\gl{(4.7)} by the subspace of solutions $\psi_\pm$ to eqs.\gl{(4.11)}
can be described in terms of sheaf cohomology groups and
Kodaira-Spencer deformation theory.  For discussion in Euclidean
signature and references see~\cite{Po}.

\bigskip
\bigskip\noindent
{\bf 4.2\ \ Transformations of tetrads on self-dual manifolds}
\smallskip

The explicit form of the transformations of a (conformal) SD
tetrad	$\{W_A, W_{\bar A}\}$ on $U\subset M$ can easily be obtained
from
formulae \gl{(4.3a)}-\gl{(4.8c)}. Namely, let us choose any solution $\psi$ of
eqs.\gl{(4.7)} and split it by formulae \gl{(4.8a)}-\gl{(4.8c)}. Then we have
\bea\label{(4.13a)}
\d_\psi W_{\bar 1} &=& \oint_{S^1}\frac{d\l}{2\pi i\l}[\psi_+,
W_{\bar 1} -\l W_2]=\oint_{S^1}\frac{d\l}{2\pi i\l}[\psi_-,
W_{\bar 1} -\l W_2] \nonumber \\
&& =[\psi^0_+, W_{\bar 1}]= [\psi^0_-, W_{\bar 1}]-[\psi^1_-, W_2],
\eea
\bea\label{(4.13b)}
\d_\psi W_{\bar 2} &=& \oint_{S^1}\frac{d\l}{2\pi i\l}[\psi_+,
W_{\bar 2} -\l W_1]=\oint_{S^1}\frac{d\l}{2\pi i\l}[\psi_-,
W_{\bar 2} -\l W_1] \nonumber \\
&& =[\psi^0_+, W_{\bar 2}]= [\psi^0_-, W_{\bar 2}]-[\psi^1_-, W_1],
\eea
\bea\label{(4.13c)}
\d_\psi W_{1} &=& \oint_{S^1}\frac{d\l}{2\pi i\l}[\psi_-,
W_{1} -\frac{1}{\l}W_{\bar 2}]=\oint_{S^1}\frac{d\l}{2\pi i\l}[\psi_+,
W_{1} -\frac{1}{\l}W_{\bar 2}] \nonumber \\
&& =[\psi^0_-, W_{1}]= [\psi^0_+, W_{1}]-[\psi^1_+, W_{\bar 2}],
\eea
\bea\label{(4.13d)}
\d_\psi W_{2} &=& \oint_{S^1}\frac{d\l}{2\pi i\l}[\psi_-,
W_{2} -\frac{1}{\l}W_{\bar 1}]=\oint_{S^1}\frac{d\l}{2\pi i\l}[\psi_+,
W_{2} -\frac{1}{\l}W_{\bar 1}] \nonumber \\
&& =[\psi^0_-, W_{2}]= [\psi^0_+, W_{2}]-[\psi^1_+, W_{\bar 1}],
\eea
where the contour $S^1=\{\l\in\C:  |\l |=1\}$ circles once around $\l =0$
and $\psi^n_\pm$ are coefficients in the Fourier series \gl{(4.8c)}.

\medskip

Let us consider solutions $\psi_\pm$ from \gl{(4.5)}-\gl{(4.8c)}
such that $\psi =
\psi_+-\psi_-=0$ on $\cp_0$. The condition $\psi_+=\psi_-$ on $\cp_0$
means that $\psi_\pm$ are the restrictions $\psi_+=\tilde\vp|_{\bar\cp_+},
\psi_-=\tilde\vp|_{\bar\cp_-}$ of a globally defined vector field
$\tilde\vp$ on
$\tilde\cp\simeq U\times\C P^1$. Such vector fields $\tilde\vp$ are the
pull-back of vector fields $\vp =\vp^\mu (x)\p_\mu\in\ $sdiff$(M)$ to
$\tilde\cp$. They have the form $\tilde\vp =\vp +\vp^\l (x,\l )\p_\l$,
where $\vp^\l(x,\l)$ are quadratic in $\l$. For the vector fields
$\tilde\vp$ we have $\vp (x):=\psi^0_+=\psi^0_-$ and from formula
\gl{(4.13a)}-\gl{(4.13d)}
we obtain
\be\label{(4.14)}
\d_\vp W_A=[\vp , W_A],\quad   \d_\vp W_{\bar A}=[\vp , W_{\bar A}].
\ee
The transformations \gl{(4.14)} coincide with the transformations \gl{(3.6)}
generated by the group SDiff($M$) of volume-preserving diffeomorphisms
of $(M,\epsilon )$.

\medskip

To see a connection with the previous ``recursion" descriptions of
symmetries of the SDG equations on complex 4-manifolds we consider
a vector field $\rho\in\ $sdiff$^\C(M)$ such that
\be\label{(4.15)}
\d^0_\rho W_{\bar A}:=[\rho , W_{\bar A}]=0,\quad
 \d^0_\rho W_{A}:=[\rho , W_{A}]\ne 0.
\ee
Then we choose a solution $\psi_\rho$ of eqs.\gl{(4.7)} on $\cp_0$ such that
\be\label{(4.16)} \psi_\rho^0=-\rho ,
\ee
where $\psi_\rho^0$ is zero component in the Fourier series \gl{(4.8a)}.
By substituting the expansion \gl{(4.8a)} of $\psi_\rho$ in a Fourier series
into formula \gl{(4.7)}, we obtain the recursion relations
$$
[W_{\bar 1},\psi^{n+1}_\rho ]= [W_{2},\psi^{n}_\rho ],
$$
\be\label{(4.17)}[W_{\bar 2},\psi^{n+1}_\rho ]= [W_{1},\psi^{n}_\rho ].
\ee
It is easy to see that if we put
\be\label{(4.18)}
\d_\rho^n W_{\bar A}:=0,\quad  \d_\rho^n W_{A}:= [W_A, \psi^n_\rho ],
\ee
then eqs.\gl{(4.17)} and formulae \gl{(4.18)}
exactly reproduce the symmetries of
the SDG equations known before (see~\cite{PBR} and references therein).

\medskip

The transformations \gl{(4.18)} can be obtained from formulae
\gl{(4.13a)}-\gl{(4.13d)}. Namely,
notice that if $\psi_\rho$ satisfies eqs.\gl{(4.7)}, then for any $n\in \Z$
the vector field
\be\label{(4.19a)}\l^{-n}\psi_\rho\ee
also satisfies eqs.\gl{(4.7)}. Then split the vector field \gl{(4.19a)}
according to formulae \gl{(4.8a)}-\gl{(4.8c)},
\be\label{(4.19b)}\l^{-n}\psi_\rho =(\l^{-n}\psi_\rho )_+-(\l^{-n}\psi_\rho )_-,
\ee
and using \gl{(4.9)} choose the gauge $(\l^{-n}\psi_\rho )^0_+=0$, i.e.
\be\label{(4.19c)}
\l^{-n}\psi_\rho =(\l^{-n}\psi_\rho )_+-(\l^{-n}\psi_\rho )_-=
(\l\psi^{n+1}_\rho +\l^2\psi^{n+2}_\rho +...)-(-\psi^n_\rho -
\l^{-1}\psi^{n-1}_\rho -...).
\ee
Then we have
\be\label{(4.19d)}(\l^{-n}\psi_\rho )_+^0=0,\quad
(\l^{-n}\psi_\rho )_-^0=-\psi^n_\rho ,
\ee and formulae \gl{(4.13a)}-\gl{(4.13d)} are reduced to formulae \gl{(4.18)}. Thus the
known symmetries \gl{(4.18)} of the SDG equations are generated by the vector
fields \gl{(4.19a)} on the space $\cp_0\simeq U\times S^1$ by formulae
\gl{(4.13a)}-\gl{(4.13d)}.
We shall return to a correspondence between the `twistor'
and `recursion relations' approaches to the description of symmetries
when considering the Plebanski equations.

\vfill\eject
\bigskip\noindent
{\bf 4.3\ \ Symmetries of the Plebanski equations}
\smallskip

Symmetries of the first Plebanski equation \gl{(3.10)} can be
obtained by substituting the explicit form \gl{(3.9a)}, \gl{(3.9b)}
of a tetrad
$\{W_A, W_{\bar A}\}$ parametrized by a real function $\Omega$
into formulae \gl{(4.7)}, \gl{(4.8a)}-\gl{(4.8c)},
\gl{(4.13a)}-\gl{(4.13d)} or \gl{(4.17)}, \gl{(4.18)},
\gl{(4.19a)}-\gl{(4.19d)}. If
we want to preserve the gauge $W_{\bar A}= \p_{\bar A}$, then we
should impose the conditions $\d_\psi W_{\bar A}=0$. {}From the
formulae of Sect.4.2 it follows that for this it is sufficient
to choose always $\psi_+^0=0$ in the splitting \gl{(4.8a)}-\gl{(4.8c)}. In
particular, let us consider divergence-free vector fields $\rho
=\rho^A\p_A$ satisfying the equations $[\p_{\bar A},\rho ]=0$
and generating the algebra that we denote by $\sdiff^{\C}_2$.
Then if we put $(\psi_\rho )^0_-=-\psi^0_\rho = \rho $, then by
formulae \gl{(4.16)}-\gl{(4.19d)} we obtain the well-known Lie algebra of
symmetries $\sdiff^{\C}_2\otimes\C[\l ,\l^{-1}]$
(cf.~\cite{PBR}).

\medskip

Notice that symmetries $\d_\psi W_A$ are connected by formulae
\gl{(4.13a)}-\gl{(4.13d)} with Lie derivatives along {\it local}
vector fields
$\psi_\pm$ on $\bar\cp_\pm$. Having the explicit form of {\it
diffeomorphism-type} transformations $\d_\psi : W_A\to \d_\psi
W_A$, one can set up a problem of finding {\it functional}
transformations $\hat\d_\psi : \Omega\mapsto \hat\d_\psi\Omega$
of the K\"ahler potential $\Omega$ of a SD metric inducing the
same change of $W_A$ from \gl{(3.12a)} as under the action by
$\d_\psi$ on $W_A$. In other words, we want to find such
$\hat\d_\psi\Omega$ that the equations
\be\label{(4.20)}
\ve^{\bar
B}_{A}\ve^{CB}\p_B\p_{\bar B}(\hat\d_\psi\Omega)\p_C = \d_\psi
W_A=[\psi^0_-, W_A]=[\psi_-^0, \ve^{\bar B}_{A}\ve^{CB}
\p_B\p_{\bar B}\Omega\p_C ] \ee are satisfied. Here
we used formulae \gl{(3.12a)} and \gl{(4.13a)}-\gl{(4.13d)}.
In general, if we use only
the metric \gl{(3.14)} and do not use a tetrad $\{W_A, W_{\bar A}\}$,
then we can consider functional transformations $\hat\d
:\Omega\mapsto\hat\d\Omega$ by themselves without considering
the transformations of a tetrad.

\medskip

Consider the transformation $\hat\d : \Omega\mapsto\hat\d\Omega$ and
substitute $\Omega +\hat\d\Omega$ into the Plebanski equation \gl{(3.10)}.
Leaving terms with $\hat\d\Omega$ of the power not higher than one
(linearization), we see that $\hat\d\Omega$ satisfy the ultrahyperbolic
wave equation
$$
g^{\mu\nu}\nabla_\mu\nabla_\nu\hat\d\Omega =\frac{1}{\sqrt{\det g}}\p_\mu
(\sqrt{\det g}g^{\mu\nu}\p_{\nu}\hat\d\Omega  )=
$$
\be\label{(4.21)}
=(\p_1\p_{\bar 2}\Omega\p_2\p_{\bar 1} + \p_2\p_{\bar 1}\Omega
\p_1\p_{\bar 2} - \p_1\p_{\bar 1}\Omega\p_2\p_{\bar 2} -
\p_2\p_{\bar 2}\Omega\p_1\p_{\bar 1})\hat\d\Omega  =0,
\ee

\noindent
where we used formulae \gl{(3.12a)} and \gl{(3.14)}. Thus, finding all
infinitesimal symmetries of the first Plebanski equation is
reduced to finding  all solutions to the wave equation in a
self-dual background. The general solution of the wave equation
on Riemannian manifolds with SD conformal structures was
described by Hitchin~\cite{Hi} in twistor terms and used by
Park~\cite{Pa} in discussion of symmetries to the first
Plebanski equation on complex 4-manifolds. In the case of
ultrahyperbolic signature (2,2), the general solution of
eq.\gl{(4.21)} can also be described in twistor terms.

\medskip

Consider real analytic functions $\phi^1_0(x,\l ), \phi^2_0(x,\l
)= \overline{\phi^1_0(x,\l )}, \phi^3_0(x,\l )=\l =\exp (i\gamma)$
satisfying eqs. \gl{(3.27a)}, \gl{(3.27b)} on $\cp_0=U\times S^1$.
Let us act on
eqs.\gl{(3.27a)} by the vector field $W_1$, on eqs.\gl{(3.27b)} by the vector
field $-\l W_2$ and sum them. Taking into account that the
vector fields $\{W_A, W_{\bar A}\}$ have the form \gl{(3.9a)}, \gl{(3.9b)}
and the
function $\Omega$ satisfies the Plebanski equation \gl{(3.10)}, we
obtain that functions $\phi^A_0(x,\l )$ satisfy the wave
equation \gl{(4.21)} on $U$ for any $\l =\phi^3_0\in S^1$.  Therefore
an arbitrary complex-valued smooth function $F(\phi^1_0,
\phi^2_0, \phi^3_0)$ of $\phi^a_0$ also satisfies eq.\gl{(4.21)}. As a
function of $\phi^a_0$ it represents a free function on the
3-dimensional
real twistor space $\ct (U)$ of $U$ introduced in Sect.3.4.

\medskip

Expanding a smooth function $F( \phi^1_0, \phi^2_0, \l )$ in a
Fourier series in $\l$, we have \be\label{(4.22)}
F( \phi^1_0(x,\l ),
\phi^2_0(x,\l ), \l )=F(x,\l )=\sum^\infty_{n=-\infty}
\l^nF^n(x).  \ee
Then general smooth complex solution of eq.\gl{(4.21)} has the form
\be\label{(4.23)}\hat\d_F\Omega = F^0(y^A, y^{\bar A})=
\oint_{S^1}\frac{d\l}{2\pi i\l} F(\phi^1_0, \phi^2_0, \l ),
\ee
where $S^1=\{\l\in\C:|\l |=1\}$. Substituting the
expansion \gl{(4.22)} into the equations $\cl^0_AF=0$ which follow from
eqs.\gl{(3.27a)}, \gl{(3.27b)}
we have the recursion relations \be\label{(4.24a)} W_{\bar
1}F^{n+1}=W_2F^n,\quad W_{\bar 2}F^{n+1}=W_1F^n,\ee
where $n\in\Z$. Using eqs.\gl{(4.24a)}, one can find $F^n$ for any $n$
by choosing $F^0$ as a `seed solution'.

\medskip

We call the map \be\label{(4.24b)}\mr:\quad F^n\mapsto F^{n+1}\ee the
{\it recursion operator}.  This operator is not unique because
of the ambiguity in the inversion of $W_{\bar 1}$ and $W_{\bar
2}$. Moreover, in the absence of boundary conditions, $F^{n}$ is
determined only up to the addition of a function
$f(y^A)+g(y^{\bar A})$, where $f$ and $g$ are arbitrary
functions of $y^A$ and $y^{\bar A}$, respectively. In terms of
the recursion operator $\mr$ we have \be\label{(4.24c)}
F^{n+1}=\mr
F^n\ee and by iterating we obtain $F^n=\mr^nF^0$.

\medskip

To see a correspondence with the symmetries known before and
described e.g. in~\cite{JLP}, let us consider a complex
divergence-free vector field $\rho =\rho^A\p_A$ satisfying the
equations $[\p_{\bar A},\rho ]=0$.  Substitute $\d^0_\rho
W_A=[\rho , W_A]$ into eqs.\gl{(4.20)} ($\psi\mapsto\psi_\rho$,
$\psi^0_-\mapsto\rho$) and take into account that
$\p_A\rho^A=0$. After short calculations we see that
\be\label{(4.25a)}\hat\d^0_\rho\Omega=\rho^A\p_A\Omega\ee satisfy
eqs.\gl{(4.20)} and define a trivial complex solution of eq.\gl{(4.21)}.  To
generate nontrivial symmetries of the Plebanski eq.\gl{(3.10)}, let us
take the complex solution \gl{(4.23)} of eq.\gl{(4.21)} and choose a function
$F\equiv F_\rho =\sum_n\l^nF^n_\rho$ such that $F^0_\rho
=\rho^A\p_A\Omega =\hat\d^0_\rho\Omega$. Functions $F^n_\rho$
for $n\ne 0$ can be found by using the recursion relations
\gl{(4.24a)}-\gl{(4.24c)}.
Then \be\label{(4.25b)}\hat\d^n_\rho\Omega :=F^n_\rho \ee satisfy
eq.\gl{(4.21)} and therefore are symmetries of the first Plebanski
equation. Formula
$$
\rho^A_{n+1}:=\ve^{BA}\p_BF^n_\rho
$$
connects functions $F^n_\rho$ with functions $\rho^A_{n+1}$ used for
description of symmetries to the first Plebanski equation
in~\cite{JLP}.  The above symmetries are in a one-to-one
correspondence with coefficients $F^n_\rho$ in the expansion of
the function $F_\rho (\phi^1_0,\phi^2_0, \l )$ in $\l$ and
generate the affine Lie algebra $\sdiff^{\C}_2\otimes \C[\l ,
\l^{-1}]$.

\medskip

Notice that the wave equation \gl{(4.21)} is invariant under the
change $y^A\leftrightarrow y^{\bar A}$. If $\eta =\eta^{\bar
A}\p_{\bar A}$ is any divergence-free vector field satisfying
$[\p_A,\eta ]=0$, then \be\label{(4.27a)}\hat\d^0_\eta\Omega :=\eta^{\bar
A}\p_{\bar A}\Omega \ee is also a solution of
eq.\gl{(4.21)}. Notice that vector fields $\rho$ from \gl{(4.25a)},
\gl{(4.25b)}  and
$\eta$ from \gl{(4.27a)} are independent, commute and generate the
Lie algebra $\sdiff^{\C}_2\oplus\sdiff^{\C}_2\simeq
w_\infty\oplus w_\infty$. Taking $F^0_\eta := \eta^{\bar
A}\p_{\bar A}\Omega$ as a starting point of the recursion
procedure \gl{(4.24a)}-\gl{(4.24c)}, we introduce $F^n_\eta$ and $F_\eta
=\sum_n\l^n F^n_\eta$. Then we obtain an infinite number of new
symmetries by putting \be\label{(4.27b)}\hat\d^n_\eta\Omega :=F^n_\eta
.\ee These symmetries form the algebra
$\sdiff^{\C}_2\otimes \C[\l ,\l^{-1}]$.

\medskip

To introduce transformations preserving the reality of the
K\"ahler potential, let us consider a function $F(\phi^1_0,
\overline{\phi^1_0}, \l )$ that is real analytic and therefore
extendable to a holomorphic function $F(\phi^1_0, \phi^2_0, \l )$
on a neighbourhood $\U_0$ of $\ct (U)$ in $\tilde\cp$, where it
satisfies
\be\label{(4.28a)}
\overline{F(\phi^1_0(x,\bar\l^{-1}) , \phi^2_0(x,\bar\l^{-1}) ,
\bar\l^{-1})}= F(\phi^1_0(x,\l ), \phi^2_0(x,\l ), \l ).
\ee
On $\cp_0$ such a function is real, i.e.
$$
\overline{F(\phi^1_0, \phi^2_0, \l)}=F(\phi^1_0, \phi^2_0, \l) ,
$$
and by formula \gl{(4.23)} it gives the general real analytic solution
$\hat\d_F\Omega$ of eq.\gl{(4.21)} since $\overline{F^0}=F^0$. Indeed,
in terms of components $F^n$ of the expansion of the function $F$
in $\l$, the reality conditions \gl{(4.28a)} have the form
$$
\overline{F^n}=F^{-n}.
$$
Hence $\overline{F^0}=F^{0}$ and formula \gl{(4.23)} gives a real solution
of eq.\gl{(4.21)}.

\medskip

Having in mind all the above, let us take as $\eta$ in formulae \gl{(4.27a)}
a vector field complex conjugate to the vector field $\rho$ from
formulae \gl{(4.25a)}, \gl{(4.25b)}, i.e. choose $\eta =\bar\rho$.
Then starting with
$F^0_{\bar\rho}=\bar\rho^{\bar A}\p_{\bar A}\Omega$ and using
eqs.\gl{(4.24a)}-\gl{(4.24c)},
one can introduce a function $F_{\bar\rho}$ conjugate to $F_\rho$
in a sense that
$$
\overline{F_\rho (x, \bar\l^{-1})}=F_{\bar \rho}(x,\l )\quad
\Leftrightarrow\quad \overline{F^n_\rho}=F^{-n}_{\bar\rho}.
$$
After this we can introduce symmetries
\be\label{(4.30a)}\Delta^n_{\rho +\bar\rho}\Omega =
(\hat\d^n_\rho +\hat\d^{-n}_{\bar\rho})\Omega =
F^n_\rho +F^{-n}_{\bar\rho}=F^n_\rho +\overline{F^n_\rho},
\ee
\be\label{(4.30b)}\Delta^n_{i(\rho -\bar\rho )}\Omega =
i(\hat\d^n_\rho - \hat\d^{-n}_{\bar\rho})\Omega =
i(F^n_\rho +F^{-n}_{\bar\rho})=i(F^n_\rho -\overline{F^n_\rho}),
\ee
preserving the reality of $\Omega$.

\medskip

Symmetries of the second Plebanski equation \gl{(3.39)} can be described
analogously. Namely, linearizing eq.\gl{(3.39)}, we obtain the wave equation
\be\label{(4.31)}
(\p_1\p_{\tilde 1} - \p_2\p_{\tilde 2} + 2\p_1\p_2\Psi\p_1\p_2 -
\p^2_2\Psi\p^2_1 -\p^2_1\Psi\p^2_2)\hat\d\Psi =0,
\ee
where we use the metric inverse to the metric \gl{(3.40b)}. Using
further the linear system \gl{(3.31)} or \gl{(3.33)}, we can introduce real
analytic functions $\chi^1_0(x,\zeta ), \chi^2_0(x,\zeta ),
\chi^3_0(x,\zeta )=\zeta$ on $\cp_0$, $\zeta\in\R\cup\infty$.
{}From formula \gl{(3.30)} for $\l =\exp(i\gamma )$ we obtain $$\zeta
=\mbox{tg}(\frac{\gamma}{2}).$$ By introducing a smooth function
$F(\chi^1_0(x,\gamma ), \chi^2_0(x,\gamma ), \gamma )$, one can
represent a general solution of eq.\gl{(4.31)} in the form
$$
\hat\d_F\Psi
=\int^{2\pi}_0\frac{d\gamma}{2\pi}F(\chi^1_0,\chi^2_0,\gamma ).
$$
For discussion of this formula in the case of the
flat space $\R^{2,2}$ see e.g.~\cite{Wo, BGPPR}. One can introduce
symmetries generated by vector fields on $U$ by formulae
analogous to \gl{(4.25a)},\gl{(4.25b)}, \gl{(4.27a)}, \gl{(4.27b)}
and \gl{(4.30a)}, \gl{(4.30b)}, that we shall not do.

\bigskip
\bigskip\noindent
{\bf 4.4\ \ Relationships between SDG symmetries}
\smallskip

We shall show here that all symmetries of the SDG equations
described in Sections 4.1-4.3 are connected with each other.

\medskip

In Sect.3.4 we mentioned that all information about a SD metric $g$
on $U$ is encoded in transition functions $f^a_{+-}$ on the space
$\cp_0$, $a=1,2,3$. Accordingly, infinitesimal variations of the SD
metric are encoded in variations $\d f^a_{+-}$ of these transition
functions. In its turn, $\d f^a_{+-}$ are nothing but
components $\psi^a$ in the expansion
$$
\psi =\psi^a\frac{\p}{\p\phi^a_+}+\psi^{\bar a}
\frac{\p}{\p\phi^{\bar a}_+}
$$
in the basis $\{ {\p}/{\p\phi^a_+},  {\p}/{\p\phi^{\bar a}_+}\}$
of a vector field $\psi$ satisfying eqs.\gl{(4.7)}, i.e.
$$
\d f^a_{+-}=\psi^a .
$$
Analogously, the holomorphic components $\psi^a_\pm$ of  vector fields
$\psi_\pm$ from formulae \gl{(4.8a)}-\gl{(4.8c)}
define infinitesimal variations
$$ \d\phi^a_\pm =\psi^a_\pm $$
of the coordinates $\phi^a_\pm$ on $\bar\cp_\pm$.

\medskip

Recall that $\phi^3_+=\l ,  \phi^3_-=\l^{-1}$ and in Sections 4.1-4.3 we
considered transformations such that $\d\phi^3_\pm =\psi^3_\pm =0\
\Rightarrow\ \psi^3=0, \d\l =0$. The imposing of these restrictions
is connected with the fact that under deformations of the space
$\tilde \cp$ the holomorphic projection
\be\label{(4.36)}\pi :\quad \tilde\cp\to\C P^1\ee
must be preserved. The choice of the gauge $\psi^3_\pm
=\psi^3=\psi^\l =0$ satisfies this condition and is always possible,
which is proved in the twistor approach (see e.g.~\cite{Pe, TW}).

\medskip

Notice that on fibres of the bundle \gl{(4.36)} one can define a closed
2-form
$\omega$, holomorphic on the space $\tilde\cp$~\cite{Pe, TW}.
In fact, the equations $d\omega =0$ are equivalent to the SDG
equations~\cite{Gi}. In the local coordinates $\phi^a_\pm$
on $\tilde\cp$ this 2-form is
$$
\omega |_{\bar\cp_+}=d\phi^1_+\wedge d\phi^2_+
\quad
\omega |_{\bar\cp_-}=d\phi^1_-\wedge d\phi^2_-.
$$
On $\cp_0$ we have
$$d\phi^1_+\wedge d\phi^2_+=\l d\phi^1_0\wedge d\phi^2_0
=\l^2 d\phi^1_-\wedge d\phi^2_-,
$$
which means the quadratic dependence of $\omega|_{\bar\cp_+}$ on
$\l$.  In other words, $\omega$ is a 2-form on $\tilde\cp$ with
values in sections of the holomorphic line bundle $\pi^*\co
(2)\to\tilde\cp$ (see e.g.~\cite{HKLR}).  By deforming $\tilde\cp$
one should
preserve $\omega$ (cf.~\cite{Pe, TW, HKLR}) and therefore vector
fields $\psi$ satisfying \gl{(4.7)} have to preserve not only a
volume 4-form $\epsilon$ on $M$, but also the 2-form $\omega$ on
$\tilde\cp$.  Consider a neighbourhood $\U_0$ of $\cp_0$, a
basis $\{\p /\p\phi^a_0, \p /\p\phi^{\bar a}_0\}$ on $\U_0$ and
expand $\psi =\psi^A \p /\p\phi^A_0+ \psi^{\bar A} \p
/\p\phi^{\bar A}_0$, where it is taken into account that
$\psi^{3}=\psi^{\bar 3}=0$.
Then we obtain \be\label{(4.39)} \cl_\psi\omega
=0\quad\Leftrightarrow\quad
\psi^A\frac{\p}{\p\phi^A_0}=\ve^{BA}\frac{\p F}{\p\phi^B_0}
\frac{\p}{\p\phi^A_0}.  \ee This means that the
vector field $\psi$ acts on functions $G(\phi^A_0,\l )$ as a
Hamiltonian vector field defined on fibres of the bundle \gl{(4.36)}.
In \gl{(4.39)} the Hamiltonian $F(\phi^A_0,\l
)=F(\phi^1_0,\overline{\phi^1_0}, \l )$ is a real analytic
function on $\cp_0$ that represents a free function on the real
twistor space $\ct (U)$ of $U$. Thus, formula \gl{(4.39)} defines the
connection between vector fields $\psi$
from \gl{(4.7)}, \gl{(4.8a)}-\gl{(4.8c)} and
functions $F$ from \gl{(4.22)}, \gl{(4.23)}.

\medskip

Notice that from formulae \gl{(4.39)} it is easy to see the algebraic
properties of the transformations $\hat\d_F\Omega$ defined by
eq.\gl{(4.23)}. Namely, $$[\hat\d_F, \hat\d_G]\Omega
=\hat\d_{\{F,G\}}\Omega , $$ where
$$\{F,G\}:=\ve^{AB}\frac{\p F}{\p\phi^A_0} \frac{\p
G}{\p\phi^B_0}$$ is the Poisson bracket on fibres of
the bundle \gl{(4.36)}.

\bigskip
\section{Self-dual gravity hierarchy}

{\bf 5.1\ \ The solution space of the SDG equations}
\smallskip

In what follows we will briefly discuss symmetries from the viewpoint
of geometry of infinite-dimensional solution spaces. {}From now on,
eqs.\gl{(3.8a)}-\gl{(3.8c)} will be called the SDG equations if a
tetrad $\{W_{\bar A}, W_A\}$ is chosen in the form  \gl{(3.9a)},
\gl{(3.9b)}.  These equations are equivalent to the first Plebanski
equation.  So, the self-dual tetrads  \gl{(3.9a)}, \gl{(3.9b)} and
self-dual metrics  \gl{(3.14)} are parametrized by a function
$\Omega$ satisfying  eqs.\gl{(3.10)}.  Consider the space of all
solutions to the SDG equations and denote it by $\cn$. Then formula
\gl{(3.14)} defines the projection \be \label{(5.1)} p:\quad
\cn\to\cm \ee of the space $\cn$ on the space $\cm$ of the self-dual
metrics describing physical degrees of freedom of the SDG
gravity.

\medskip

Let us consider transformations of the space $\cn$ into itself
defined by the formula \be \Omega\ \mapsto\ \Omega'=\Omega + f{(y^A)}
+ g{(y^{\bar A})}, \label{(5.2)} \ee where $f$ and $g$ are arbitrary
functions of $y^A$ and $y^{\bar A}$, respectively. It is easy to see
that $\Omega$ and $\Omega'$ determine the same tetrad  \gl{(3.9a)},
\gl{(3.9b)} and metric  \gl{(3.14)} since functions
$f{(y^A)}+g{(y^{\bar A})}$ belong to the kernel of the projection
\gl{(5.1)}.  It should be stressed that we consider the space $\cn$
of {\it local} solutions to the SDG equations.

\medskip

In Sections 4.1-4.4 we have described solutions $\delta\Omega$ of the
linearized  (around $\Omega$) SDG equations. {}From the geometric
point of view, solutions of the linearized SDG equations are vector
fields on the solution space $\cn$, i.e. sections of the tangent
bundle $T\cn$. Therefore infinitesimal symmetries of the SDG
equations form the algebra $\Vect{(\cn)}$ of vector fields on $\cn$.
As it has been shown above, this algebra is isomorphic to the algebra
of vector fields on the real twistor space $\ct{(U)}$.

\medskip

Notice that describing symmetry algebras of the SDG equations
generated by algebras of vector fields on $U\subset M$ is equivalent
to defining homomorphisms of these algebras into the algebra
$\Vect{(\cn)}$. For instance, in Sections 4.2 and 4.3 we have
actually described the homomorphism of the algebra
$\sdiff^{\C}_2\otimes\C[\l, \l^{-1}]$ into the algebra
$\Vect^{\C}{(\cn)}$. In other words, by considering some given group
$\cg$  (e.g. the loop group $\LSDiff^{\C}_2= C^\infty (S^1,
\SDiff^{\C}_2)$) as a candidate for the symmetry group of the SDG
equations, one should say about defining  its {\it action } on the
solution space $\cn$. Accordingly, one can consider the {\it
infinitesimal action} of the group $\cg$ on $\cn$ and a {\it
homomorphism} of its Lie algebra into the algebra $\Vect {(\cn)}$ of
vector fields on $\cn$.

\bigskip
\bigskip\noindent
{\bf 5.2\ \ Abelian symmetries and flows on the solution space}
\smallskip

We consider  eqs.\gl{(3.8a)}-\gl{(3.8c)} with a tetrad of the form
\gl{(3.9a)}, \gl{(3.9b)}, parametrized by a function $\Omega$. In
Sections 4.2-4.4 we discussed symmetries of these equations generated
by vector fields $\rho =\rho^A\p_A\in\sdiff^{\C}_2
\subset\sdiff^{\C}{(M)}$ such that $[\p_{\bar A},\rho ]=0$ and
$\p_A\rho^A=0$. Consider the obtained symmetry algebra $\sdiff^{\C}_2
\otimes\C[\l,\l^{-1}]$ and the abelian subalgebra $\C^2\otimes\C[\l]$
in it, generated by the translations along $\p_A\equiv \p /\p y^A,
A=1,2$.  In other words, let us choose $\p_A$ as $\rho$ and
substitute it in  \gl{(4.15)}, \be \delta^0_A W_B :=\p_A W_B=[W_B,
\psi^0_A]\quad \Rightarrow\quad \psi^0_A=\ve^{\bar B}_A\p_{\bar
B}\psi , \label{(5.3a)} \ee where $\psi =\p_1\phi\p_2-\p_2\phi\p_1,
\phi =\Omega -{(y^1y^{\bar 1} -y^2y^{\bar 2})}$.  Then with the help
of recurrence formulae {(4.25)} we shall find all $\psi^n_A$ with
$n>0$ and substitute them into formulae  \gl{(4.18)}, \be \delta^n_A
W_{\bar B}=0,\quad \delta^n_A W_{B}=[W_B, \psi^n_A], \label{(5.3b)}
\ee assuming that $\psi^n_A=0$ for $n<0$. Accordingly, in formula
\gl{(4.25a)} we set \be F^0_A=\hat\delta^0_A\Omega :=\p_A\Omega ,
\label{(5.4a)} \ee and find all \be F^n_A= \hat\delta^n_A\Omega
\label{(5.4b)} \ee with $n>0$ using the recurrence relations
\gl{(4.24a)}-\gl{(4.24c)}  (see Sections 4.2 and 4.3).  Formulae
\gl{(5.3a)}, \gl{(5.3b)} and  \gl{(5.4a)},  \gl{(5.4b)} define the
action of the algebra $\C^2\otimes \C[\l ]$ on $\{W_{\bar B}, W_B\}$
and $\Omega$.

\medskip

In Sect.5.1 we discussed  infinitesimal symmetries of the SDG
equations as vector fields on the solution space $\cn$. In
particular, an infinite number of commuting symmetries  \gl{(5.4a)},
 \gl{(5.4b)} corresponds to an infinite number of commuting vector
fields $\hat\delta^n_A$ on $\cn$ with the components $F^n_A=
\hat\delta^n_A\Omega$ at the point $\Omega\in\cn$, $n=0,1,...\ $.  To
these vector fields we can correspond the system of differential
equations \be \frac{\p}{\p t^A_n}\Omega =F^n_A \label{(5.5a)} \ee on
$\cn$, where $t^A_n\in \C$ are complex parameters, $A=1,2,\
n=0,1,...\ $.  Let us also introduce the parameters $t^A_{-1}\in\C$
by the formula \be \frac{\p}{\p t^A_{-1}}\Omega =\varepsilon^{\bar
B}_A \frac{\p}{\p y^{\bar B}}\Omega , \label{(5.5b)} \ee where
$\varepsilon^{\bar B}_A  $ are given by formulae  \gl{(3.11)}.  The
dependence of $\Omega$ on $t={(t^A_n)}$ can be recovered by solving
successively eqs.\gl{(5.5a)}, \gl{(5.5b)}.  Solutions $\Omega
(t)=\Omega (y^{\bar A} + \varepsilon^{\bar A}_B t^B_{-1}, y^A+t^A_0,
t^A_1, ...)$ of the dynamical systems \gl{(5.5a)}, \gl{(5.5b)} are
called {\it integral curves} or {\it flows} of vector fields
$\hat\delta^n_A$. For small $t^A_n$'s we have \be \Omega {(t)}=\Omega
{(t=0)}+\sum_{A,n}t^A_nF^n_A +O{(t)}.  \label{(5.6)} \ee


We can represent $\Omega$ as a function of the parameters $t^A_n$
with $n\ge -1$ alone, with dependence on the coordinates $y^{\bar A},
y^A$ recovered by substituting $y^{\bar 1}+t^2_{-1}$ for  $t^2_{-1}$,
and so on. Recall that  solutions of the linearized SDG equations are
vector fields tangent to the flows and  eqs.\gl{(5.5a)}, \gl{(5.5b)}
are mutually consistent because the flows commute.  We denote by
$\Gamma_+$ the group corresponding to the algebra $\C^2\otimes\C [\l
]$ with the generators $\{\p^n_A\}$.  The parameters $t^A_n$ can be
considered as coordinates on a subspace $\T$ of the solution space
$\cn$. Notice that the space $\cn$ of local solutions to the SDG
equations can be described as an orbit of the loop group
$\LSDiff^{\C}_2$, and the coordinates $t={(t^A_n)}={(t^A_{-1}, t^A_0,
t^A_1, ... )}$ parametrize an orbit $\T$ of the abelian subgroup
$\Gamma_+$ of the group $\LSDiff^\C_2$. The action of this group on a
seed solution $\Omega  {(t=0)}$ embeds it in a family of new
solutions $\Omega  {(t)}$, labelled by the parameters $t ={(t^A_n)}$.
To sum up, the flows are generated by the translations along the
vector fields $\p^n_A={\p}/{\p t^A_n}$ on the orbit $\T$ of the group
$\Gamma_+$ acting on the solution space $\cn$.

\bigskip
\bigskip\noindent
{\bf 5.3\ \ Higher flows and SDG hierarchy}
\smallskip

Let us move on to more concrete description of flows on the solution
space $\cn$ of the SDG equations. First of all, using equations
\gl{(4.17)}--{(4.30)}, we show how $\delta^n_A$ are
connected with $\hat\delta^n_A$. We have \be \delta^n_A
W_1=\p_1\p_{\bar 2}\hat\delta^n_A\Omega\p_2 - \p_2\p_{\bar
2}\hat\delta^n_A\Omega\p_1= \p_1\p_{\bar 2}\p^n_A\Omega\p_2 -
\p_2\p_{\bar 2}\p^n_A\Omega\p_1=\p^n_AW_1, \label{(5.7a)} \ee \be
\delta^n_A W_2=\p_1\p_{\bar 1}\hat\delta^n_A\Omega\p_2 - \p_2\p_{\bar
1}\hat\delta^n_A\Omega\p_1= \p_1\p_{\bar 1}\p^n_A\Omega\p_2 -
\p_2\p_{\bar 1}\p^n_A\Omega\p_1=\p^n_AW_2.  \label{(5.7b)} \ee We see
that $\delta^n_AW_B=\p^n_AW_B$, $\hat\delta^n_A\Omega =\p^n_A\Omega$
and therefore we can identify $\delta^n_A$ with $\hat\delta^n_A$ and
do not write a hat over $\delta$.  On the other hand, from
eqs.\gl{(4.17)} we have \be
\p^n_AW_1=\d^n_AW_1=[W_1,\psi^n_A]=\p_{\bar 2}\psi^{n+1}_A,
\label{(5.8a)} \ee \be \p^n_AW_2=\d^n_AW_2=[W_2,\psi^n_A]=\p_{\bar
1}\psi^{n+1}_A, \label{(5.8b)} \ee and substituting  \gl{(5.7a)},
\gl{(5.7b)} into \gl{(5.8a)}, \gl{(5.8b)}, one can show that \be
\psi^{n+1}_A=\p^n_A\psi , \label{(5.9a)} \ee where \be \psi
:=\p_1\phi\p_2 -\p_2\phi\p_1, \label{(5.9b)} \ee \be \phi :=\Omega
-\Omega_0=\Omega -  (y^1y^{\bar 1}-y^2y^{\bar 2}).  \label{(5.9c)}
\ee The function $\phi$ is often used instead of $\Omega$ by
considering the first Plebanski equation.

\medskip

Now we should take into account that by definition $\p_A\Omega
=\p^0_A\Omega$, $\p_{\bar A}\Omega =\ve^B_{\bar A} \p^{-1}_B\Omega$.
Then we have \be W_{\bar 1}=\p_{\bar 1}=\p^{-1}_2, \quad W_{\bar
2}=\p_{\bar 2}=\p^{-1}_1, \label{(5.10a)} \ee \be
W_A=\p_A^0+\psi^0_A, \quad \psi^0_A=\p^0_A\psi , \label{(5.10b)} \ee
when acting on $\phi^a_+$. Let us introduce the operators \be
\cl^n_A:=\p^n_A-\l  (\p^{n+1}_A +\psi^{n+1}_A), \label{(5.11)} \ee
with $n=-1,0,1,...$, which for $n=-1$ coincide with the operators
$\cl^+_A$ from  eqs.\gl{(3.15)}. Then we consider the linear
equations \be\cl^n_A\phi^a_+=0, \label{(5.12)}\ee where $\phi^a_+$
are complex coordinates on $\bar\cp_+\subset\tilde\cp$.  The
compatibility conditions of  eqs.\gl{(5.12)} have the form
\be[\cl^m_A,\cl^n_B]=0\quad\Leftrightarrow\quad [\p^m_A-\l
(\p^{m+1}_A+ \psi^{m+1}_A), \p^n_B-\l  (\p^{n+1}_B+\psi^{n+1}_B)]=0.
\label{(5.13a)} \ee These equations can be rewritten as follows:
\be\p^m_A\psi^{n+1}_B-\p^n_B\psi^{m+1}_A =0, \label{(5.13b)}\ee
\be\p^{m+1}_A\psi^{n+1}_B-\p^{n+1}_B\psi^{m+1}_A +[\psi^{m+1}_A,
\psi^{n+1}_B]=0,\label{(5.13c)}\ee where $m,n=-1, 0, 1, ...\ $.

\medskip

Put $m=n=-1$ in  eqs.\gl{(5.13a)}-\gl{(5.13c)} and use
\gl{(5.9a)}-\gl{(5.9c)} and  \gl{(5.10a)}, \gl{(5.10b)}.  Then
eqs.\gl{(5.13a)}-\gl{(5.13c)} coincide with the SDG equations which
are reduced to the first Plebanski equation in the coordinates
$t^A_{-1}, t^A_0$:  \be\p^0_1\p^{-1}_2\phi -\p^0_2\p^{-1}_1\phi
+\p^0_1\p^{-1}_1\phi \p^0_2\p^{-1}_2\phi -
\p^0_1\p^{-1}_2\phi\p^0_2\p^{-1}_1\phi =0.  \label{(5.14a)} \ee It is
easy to see that for $m=-1, n\ge 0$  eqs.\gl{(5.13a)}-\gl{(5.13c)}
coincide with eqs.\gl{(5.8a)}, \gl{(5.8b)} defining symmetries of the
 SDG equations. Using \gl{(5.9a)}-\gl{(5.9c)}, these equations can be
rewritten as equations on the function $\phi$, \be \p^0_1\p^n_B\phi
-\p^{-1}_1\p^{n+1}_B\phi +\p^{-1}_1\p^0_1\phi \p^0_2\p^n_B\phi -
\p^{-1}_1\p^0_2\phi\p^0_1\p^n_B\phi =0, \label{(5.14b)}\ee \be
\p^0_2\p^n_B\phi -\p^{-1}_2\p^{n+1}_B\phi +\p^{-1}_2\p^0_1\phi
\p^0_2\p^n_B\phi - \p^{-1}_2\p^0_2\phi\p^0_1\p^n_B\phi =0.
\label{(5.14c)}\ee If we differentiate  eqs.\gl{(5.14b)} w.r.t.
$t^2_{-1}$ and  eqs.\gl{(5.14c)} w.r.t.  $t^1_{-1}$ and subtract the
second equation from the first equation, then we obtain the
linearized first Plebanski equation with $\d^n_A\phi =\p^n_A\phi$.
Thus, it follows from  eqs.\gl{(5.14b)},  \gl{(5.14c)} that
$\p^n_A\phi$ satisfies to the linearized  (around $\phi$) SDG
equations. Finally, for $m\ge 0$, $n\ge 0$
eqs.\gl{(5.13a)}-\gl{(5.13c)} coincide with commutativity conditions
for flows generated by symmetries $\d^n_A{(=\p^n_A)}$ on the solution
space $\cn$ of the SDG equations.

\medskip

To rewrite  eqs.\gl{(5.13a)}-\gl{(5.13c)} in terms of the function
$\phi$ for all $m,n\ge -1$, let us substitute the expression
\gl{(5.9a)} for $\psi^{n+1}_A$ into eqs.\gl{(5.13a)}-\gl{(5.13c)}.
Then we have \be \p_A^{m+1}\p_B^n\phi - \p_A^{m}\p_B^{n+1}\phi +
\p_A^{m}\p_1^{0}\phi \p_B^{n}\p_2^{0}\phi   - \p_A^{m}\p_2^{0}\phi
\p_B^{n}\p_1^{0}\phi =0, \label{(5.15)}\ee where $m,n=-1,0,1,...\ $.
The main conclusion is that {\it the function $\phi$ depends on an
infinite number of variables} $t={(t^A_n)}$, four of which can be
identified with the coordinates on $U$: $t^A_{-1}=\ve^A_{\bar
B}y^{\bar B}, t^A_0=y^A.$ Variables $t^A_n$, $n\ge -1$, parametrize
an orbit $\T$ of the abelian symmetry group $\Gamma_+\subset
\LSDiff^{\C}_2$ in the solution space $\cn$ of the SDG equations.
This interpretation of abelian symmetries and higher flows, according
to which integrable equations are described as dynamical systems on
an infinite-dimensional Grassmannian manifolds associated with loop
groups, is standard in the theory of integrable systems  (see
e.g.~\cite{Sa, DJKM, SW, Ne}).  So solutions of any integrable
equation depend on an infinite number of ``times" and obey an
infinite set of differential equations - {\it hierarchy} - generated
by flows along the higher ``times"  (higher flows).

\medskip

We shall call  eqs.\gl{(5.15)} the {\it SDG hierarchy generated by
recursion from the translations along $\p_1=\p_1^0$ and}
$\p_2=\p_2^0$. The vector fields $\p_A^n$ labelled by consecutive
values of $n$ are related by the recursion operator $\mr$:
$\p_A^{n+1}=\mr\p^n_A\ \Rightarrow\ \p^n_A=\mr^n\p^0_A$ {(see
Sect.4.3 for the definition of $\mr$)}.  One can also introduce an
SDG hierarchy generated by inverse recursion from the translations
$\p_{\bar 1}$ and $\p_{\bar 2}$. To obtain these equations we
introduce an infinite number of complex parameters $t^{\bar A}_{-n}$,
$n=-1, 0, 1,...$, so that $\p^1_{\bar A}\Omega = \ve^B_{\bar
A}\p_B\Omega$, $\p^0_{\bar A}\Omega=\p_{\bar A}\Omega$.  We also
introduce an infinite number of linear differential operators \be
\cl^{-n}_{\bar A}:=\p^{-n}_{\bar A}-\frac{1}{\l}{(\p^{-n-1}_{\bar A}+
\psi^{-n-1}_{\bar A})}, \label{(5.16)} \ee where $\psi^{-n-1}_{\bar
A}$ are some vector fields on $U$, and consider linear differential
equations \be \cl^{-n}_{\bar A}\phi^a_-=0, \label{(5.17)} \ee where
$\phi^a_-$ are complex coordinates on $\bar\cp_-\subset\tilde\cp$,
$a=1,2,3$, $n=-1,0,1,2,...\ $.  Notice that  eqs.\gl{(5.17)} with
$n=-1$ can be obtained from eqs.\gl{(3.15)} by the action of the
 operator $\tau$ of real structure  (see Sect.3.4), i.e. by the
combination of map $\l\mapsto\bar\l^{-1}$ and complex conjugation.

\medskip

The compatibility conditions of  eqs.\gl{(5.17)} are the equations
\be [\cl^{-m}_{\bar A}, \cl^{-n}_{\bar B}]=0\quad
\Leftrightarrow\quad [\p^{-m}_{\bar A}-\frac{1}{\l}{(\p^{-m-1}_{\bar
A}+\psi^{-m-1}_{\bar A})}, \p^{-n}_{\bar
B}-\frac{1}{\l}{(\p^{-n-1}_{\bar B}+\psi^{-n-1}_{\bar B})}]=0,
\label{(5.18a)} \ee which are equivalent to the equations \be
\p^{-m}_{\bar A}\psi^{-n-1}_{\bar B} - \p^{-n}_{\bar B}
\psi^{-m-1}_{\bar A}=0, \label{(5.18b)}\ee \be \p^{-m-1}_{\bar
A}\psi^{-n-1}_{\bar B} - \p^{-n-1}_{\bar B} \psi^{-m-1}_{\bar
A}+[\psi^{-m-1}_{\bar A},\psi^{-n-1}_{\bar B}] =0.
\label{(5.18c)}\ee It is not difficult to verify that for $m=n=-1$
eqs.\gl{(5.18a)}-\gl{(5.18c)} are equivalent to the first Plebanski
equation, since $\psi^0_{\bar A}= \p^0_{\bar A}\bar\psi$, $\bar\psi =
\p^0_{\bar 1}\phi\p^0_{\bar 2}- \p^0_{\bar 2}\phi\p^0_{\bar 1}$. For
$m=-1, n\ge 0$  eqs.\gl{(5.18a)}-\gl{(5.18c)} are reduced to the
equations \be \p^{-n}_{\bar A}\overline{W_B}=\d^{-n}_{\bar
A}\overline{W_B} = [\overline{W_B},\psi^{-n}_{\bar A}]=\ve^A_{\bar
B}\p_A\psi^{-n-1}_{\bar A}, \label{(5.19)}\ee defining symmetries of
the SDG equations. These symmetries form the algebra
$\C^2\otimes\C[\l^{-1}]$ with the generators $\d^{-n}_{\bar A}
(=\p^{-n}_{\bar A}), n\ge -1$. In  \gl{(5.19)} $\overline{W_B}$ is
complex conjugate to $W_B$.

\medskip

One can easily obtain from  eqs.\gl{(5.19)} and definitions
\gl{(3.9a)}, \gl{(3.9b)}, that \be \psi^{-n}_{\bar A}=\p^{-n-1}_{\bar
A}\bar\psi , \label{(5.20)}\ee where $\bar\psi$ is the vector field
complex conjugate to the vector field  \gl{(5.9b)}. By substituting
\gl{(5.20)} into \gl{(5.18a)}-\gl{(5.18c)}, we obtain the equations
\be \p^{-m-1}_{\bar A}\p^{-n}_{\bar B}\phi - \p^{-m}_{\bar
A}\p^{-n-1}_{\bar B}\phi + \p^{-m}_{\bar A}\p^0_{\bar 1}\phi
\p^{-n}_{\bar B}\p^0_{\bar 2}\phi - \p^{-m}_{\bar A}\p^0_{\bar 2}\phi
\p^{-n}_{\bar B}\p^0_{\bar 1}\phi =0, \label{(5.21)}\ee where $m,n\ge
-1$. Now recall that imposing the reality conditions on symmetry
transformations has led us in Sect.4.3 to the conditions
$\d^{-n}_{\bar A}\Omega =\overline{\d^{n}_{A}}\Omega $, where
$\Omega$ is connected with $\phi$ by formula  \gl{(5.9c)}. But
$\d^{-n}_{\bar A}\Omega = \p^{-n}_{\bar A}\Omega
=\overline{\d^{n}_{A}}\Omega =\overline{\p^{n}_{A}} \Omega$ and
therefore we obtain that \be \p^{-n}_{\bar A}=\overline{ \p^{n}_{A}}
. \label{(5.22)}\ee So eqs.\gl{(5.21)} of the SDG hierarchy generated
 by $\p_{\bar A}$ are complex conjugation of  eqs.\gl{(5.15)} and
therefore do not contain new degrees of freedom and do not give any
additional information in comparison with  eqs.\gl{(5.15)}. As such,
we shall consider only  eqs.\gl{(5.15)} and call them the {\it SDG
hierarchy}.

\bigskip
\bigskip\noindent
{\bf 5.4\ \ Twistor geometry and truncated SDG hierarchy}
\smallskip

The flows generated by $\p^{-1}_A$ and $\p^0_A$ are space-time
translations $$\phi  (t^A_{-1}, t^A_0, ...)\mapsto\phi
{(t^A_{-1}+\a^A_{-1}, t^A_0+\a^A_{0}, ...)}$$ and therefore give
rather trivial new solutions. But the higher flows generated by
$\p^n_A$ with $n\ge 1$ give less trivial new solutions.  We have
shown above that the SDG equations are embedded in an infinite system
of partial differential equations, and every solution of the SDG
equations can be extended to a solution of the infinite system of
equations  \gl{(5.15)} of the SDG hierarchy.

\medskip

To understand the geometric meaning of these equations, let us
truncate the SDG hierarchy, i.e. suppose that \be\p^n_A\phi
=0\quad\mbox{for}\quad n>r,  \label{(5.23)}\ee where $r\ge 0$ is an
integer. Then we have a finite number of $2{(r+1)}$ linear equations
\gl{(5.12)} with $-1\le n\le r-1$ on complex functions $\phi^a_+$ on
$\bar\cp_+\subset\tilde\cp$ and a finite system of eqs.\gl{(5.15)}
 with $m,n=-1,0,..., r-1$ on a function $\phi$ of $2{(r+2)}$
variables $t={(t^A_{-1}, t^A_0,..., t^A_r)}$. This finite system of
equations on $\phi$ we shall call the {\it SDG hierarchy up to level
r}, and denote it by SDG{($r$)}. In particular, SDG{(0)} equations
coincide with the first Plebanski equation, and the SDG hierarchy
equations are obtained when $r\to\infty$ and will be denoted by
SDG{($\infty$)}.

\medskip

The parameters $t^A_n$ with $n=-1, 0, ..., r$ are local coordinates
on some complex manifold of dimension $2{(r+2)}$ which we denote by
$M_r$. In the special case $r=0$, the space $M_0$ is a self-dual
4-manifold described locally by the Plebanski equations. Manifolds
$M_r$ and their geometry have been described by Alekseevsky and
Graev~\cite{AG} in the differential geometry setting. We shall
briefly describe their approach and compare it with our approach
based on ideas of symmetry, integrability etc.

\medskip

Recall that any holomorphic line bundle over the projective line $\C
P^1$ is isomorphic to the holomorphic line bundle $\co {(r+1)}\to\C
P^1$ of Chern class ${(r+1)}$ for some integer $r$. For $r\ge -1$ the
space of holomorphic sections of $\co  {(r+1)}$ is $H^0{(\C P^1, \co
{(r+1)})}\simeq S^{r+1}\C^2=\C^{r+2}$. The space $\C^{r+2}$ is the
space of irreducible representation of type ${(\frac{1}{2}{(r+1)},
0)}$ of the group $SL{(2,\C)}$ having noncompact real forms
$SU{(1,1)}\simeq SL{(2,\R)}\simeq Sp{(2, \R)}$.  A holomorphic vector
bundle $N$ of rank $l$ over the Riemann sphere $\C P^1$ is called
${(r+1)}$-{\it isotonic} if it is decomposed into a direct sum of $l$
copies of the line bundle $\co  {(r+1)}: N\simeq l\co  {(r+1)}= \co
{(r+1)}\oplus ...\oplus  \co {(r+1)}$~\cite{AG}. We shall consider
the case $l=2$, i.e. the bundle $N=\co  {(r+1)}\oplus\co  {(r+1)}$.
The space of holomorphic sections $H^0{(\C P^1, N)}$ of this bundle
$N$ carries a tensor product structure  {(Grassmann structure)},
since \be H^0{(\C P^1,N)}=H^0{(\C P^1, N\otimes \co{(-r-1)})}\otimes
H^0{(\C P^1, \co {(r+1)})}\simeq \C^2\otimes \C^{r+2}.
\label{(5.24)} \ee On the space $\C^2$ we always define the
antisymmetric bilinear form with components $\ve_{AB}$  (a volume
form). Therefore the automorphism group of the bundle $N=\co
(r+1)\oplus\co {(r+1)}$ is the group $SL{(2,\C)}\otimes SL{(2,\C)}$,
where the left group $SL{(2,\C)}$ acts irreducibly on $\C^{2}$ and
the right group $SL{(2,\C)}$ acts irreducibly on $\C^{r+2}$.

\medskip

Consider a complex 3-dimensional manifold $\cz$. Let $\pi :  \cz\to\C
P^1$ be a holomorphic fibration over the Riemann sphere $\C P^1$. A
rational curve $P$ in $\cz$ is called ${(r+1)}$-{\it isotonic} if its
normal bundle $N_P=T\cz /TP$ is isomorphic to the bundle $\co
{(r+1)}\oplus\co  {(r+1)}$ over $\C P^1\hookrightarrow\cz$~\cite{AG}.
For such curves the space $H^0{(P, N_P)}$ of holomorphic sections of
$N_P$ is isomorphic to the space $E_P\otimes H_P\simeq
\C^2\otimes\C^{r+2}$. A section $\sigma :  \C P^1\to\cz$ is called
${(r+1)}$-{\it isotonic} if the curve $\sigma  (\C P^1)$ is
${(r+1)}$-isotonic~\cite{AG}. At last, let us consider the bundle $L=
\co  (2r+2)$ over $\C P^1\hookrightarrow\cz$ and its pull-back
$\pi^*L$ to $\cz$.  Denote by $\Lambda^2_\pi$ the sheaf over $\cz$ of
holomorphic 2-forms on fibres of $\pi$ and put $\Lambda^2_\pi
{(L)}:=\Lambda^2_\pi\otimes\pi^*L$. A section $\omega\in H^0{(\cz ,
\Lambda^2_\pi  {(L)})}$ of the sheaf $\Lambda^2_\pi  {(L)}$ is called
a {\it relative symplectic structure} of type $L$ on $\cz$ if it is
closed and nondegenerate on fibres~\cite{AG}. The integer deg$L=
c_1{(L)}=2{(r+1)}$ is called the weight of the structure $\omega$.

\medskip

{\bf Theorem.} {\it Let $\cz$ be a complex 3-manifold such that

{(i)} $\cz$ is a holomorphic fibre bundle $\pi : \cz\to\C P^1$ over
$\C P^1$,

{(ii)} the bundle admits a  {(r+1)}-isotonic section $\sigma$ of
$\pi$,

{(iii)} there exists a holomorphic relative symplectic 2-form
       $\omega$ of weight $2{(r+1)}$ on $\cz$.

\noindent Then there exists a family of ${(r+1)}$-isotonic sections
containing $\sigma$ and the parameter space of holomorphic sections
is a $2{(r+2)}$-dimensional complex manifold $M_r$ with a natural
$SL{(2,\C)}\otimes 1_{r+2}$-structure.}

\medskip

For proof see~\cite{AG}. Notice only that the proof is based on a
theorem of Kodaira~\cite{Ko} which asserts that if the sheaf
cohomology group $H^1{(\sigma  {(\C P^1)}, N_\sigma )}$ vanishes then
a sufficiently small deformation of the ${(r+1)}$-isotonic section
$\sigma$ remains a ${(r+1)}$-isotonic section and may be integrated
to a deformation of the projective line $\sigma  {(\C P^1)}$, which
makes the parameter space of all holomorphic sections of $\cz$ a
complex manifold $M_r $ with the tangent space $T_{\sigma'} M_r$ at
the point $\sigma'\in M_r$ isomorphic to \be H^0{(\sigma'  {(\C
P^1)}, N_{\sigma'} )}\simeq E_{\sigma'}\otimes H_{\sigma'}
\simeq\C^2\otimes\C^{r+2}.  \label{(5.25)} \ee Therefore the tangent
bundle $TM_r$ is isomorphic to the tensor product $E\otimes H$ of a
holomorphic vector bundle $E\to M_r$ of rank 2 with fibres
$E_{\sigma'}\simeq\C^2$ at $\sigma'\in M_r$ and the {\it trivial}
vector bundle $H=M_r\times\C^{r+2}$.  The holonomy group
$SL{(2,\C)}\simeq SL{(2,\C)}\otimes 1_{r+2}$ of the manifold $M_r$ is
very ``small", which impose severe restrictions on geometry of $M_r$.
In particular, a metric on $M_r$ is parametrized by the function
$\phi  {(t^A_{-1}, t^A_0, ..., t^A_r)}$, and any metric connection
always has a torsion except the case $r=0$, when one can introduce
the standard Levi-Civita connection. Notice that ${(r+1)}$-isotonic
sections, that are real  {(i.e. preserved by $\tau$)}, are
parametrized by a real $2{(r+2)}$-dimensional submanifold of $M_r$.

\medskip

How are the results by Alekseevsky and Graev connected with ours?
What is the geometric meaning of linear equations  \gl{(5.12)} and
the SDG{($r$)} equations? Answers are the following.  As a manifold
$\cz$ we considered the 3-dimensional complex manifold $\tilde\cp
=\cp_+\cup\cp_0\cup\cp_-$ that is diffeomorphic to the manifold
$U\times S^2$ as a real manifold  {(see Sect.3.4)}.  This manifold
$\tilde\cp $ is a holomorphic fibre bundle \be \pi : \tilde\cp \to\C
P^1 \label{(5.26)}\ee over the projective line $\C P^1$  {(see
Sect.4.4)}. Sections $\sigma$ of the bundle  \gl{(5.26)} over $\bar
H_+\subset\C P^1$ are defined by the functions $\phi^A_+,
\sigma{(\bar H_+)}={(\l ,\phi^A_+ )}$, and over $\bar H_-\subset\C
P^1$ by the functions $\phi^A_-, \sigma{(\bar H_-)}=  {(\l^{-1}
,\phi^A_-)}$, where $\phi^A_\pm$ were introduced in Sect.3.3.  The
relative symplectic 2-form $\omega$ in the local coordinates
$\phi^A_\pm$ on $\bar\cp_\pm$ has the form \be \omega
|_{\bar\cp_+}=d\phi^1_+\wedge d\phi^2_+, \quad \omega
|_{\bar\cp_-}=d\phi^1_-\wedge d\phi^2_-, \label{(5.27)} \ee and was
introduced in Sect.4.4.

\medskip

Equations  \gl{(5.12)} with $n=-1, 0, ..., r-1$ define
${(r+1)}$-isotonic sections $\sigma  {(\bar H_+)} = $ $(\l ,$
$\phi^A_+(t^B_{-1}, ...,$ $t^B_r, \l ))$ of the bundle  \gl{(5.26)}.
These sections exists if  the compatibility conditions
\gl{(5.13a)}-\gl{(5.13c)} are satisfied. Equations
\gl{(5.13a)}-\gl{(5.13c)} are equivalent to  eqs.\gl{(5.15)} with
$-1\le m,n\le r-1$  {(SDG{($r$)} equations)}.  Notice that if
$\phi^A_\pm$ define a ${(r+1)}$-isotonic section of the bundle
\gl{(5.26)}, the relative symplectic form $\omega$ takes values in
the bundle $\pi^*\co  {(2r+2)}$ and therefore on $\cp_0=\bar\cp_+
\cap\bar\cp_-$ we have \be d\phi^1_+\wedge d\phi^2_+= \l^{2r+2}
d\phi^1_-\wedge d\phi^2_-.  \label{(5.28)}\ee This means that $\omega
|_{\bar\cp_+}$ may be written as a polynomial of degree $2{(r+1)}$ in
$\l\in\bar H_+$.

\medskip

In the ``flat" limit, when $\phi =0$ and the SDG{($r$)} equations are
satisfied identically, solutions of equations  \gl{(5.12)} and
\gl{(5.17)} have the form \be
\phi^A_+=\sum_{n=-1}^r\l^{n+1}t^A_{r-n-1}, \quad
\phi^A_-=\sum_{n=-1}^r\l^{n-r}t^A_{r-n-1}.  \label{(5.29a)}\ee This
means that in the ``flat" limit the space $\tilde\cp$ is
biholomorphic to the bundle $\co  {(r+1)}\oplus\co {(r+1)}\to\C P^1$,
and the functions \gl{(5.29a)} define global sections of this bundle.
To clarify the connection with representations of the groups $SL{(2,\C)}$
and $SU{(1,1)}$, we introduce on $\C P^1$ homogeneous coordinates
${(\l^A)}={(\l^1,\l^2)}$ so that $\l = \l^2/\l^1$. We can then write
the coordinates $t^A_n$ as $t^A_{A_1...A_{r+1}}$, symmetric over
indices $A_1...A_{r+1}$, related to the coordinates  \gl{(5.29a)} by
\be\sum_{A_1,...,A_{r+1}}t^A_{{A_1...A_{r+1}}}\l^{A_1}...\l^{A_{r+1}}=
{(\l^1)}^{r+1}\sum_{n=-1}^r\l^{n+1}t^A_{n,r-n}=
{(\l^1)}^{r+1}\sum_{n=-1}^r\l^{n+1}t^A_{r-n-1}.  \label{(5.29b)}\ee
Here we have introduced $t^A_{n,r-n}:=t^A_{r-n-1}$.

\medskip

The SDG{($r$)} equations always have a family of solutions containing
the trivial solution $\phi =0$ since due to the Kodaira
theorem~\cite{Ko} there exist a family of ${(r+1)}$-isotonic sections
of the bundle  \gl{(5.26)} containing the sections  \gl{(5.29a)},
\gl{(5.29b)}. In particular,  eqs.\gl{(3.15)} define $1$-isotonic
sections of the bundle  \gl{(5.26)},  and such sections exist if
$\phi$ satisfies the first Plebanski equation. The existence of
solutions $\phi$ of  eqs.\gl{(5.15)} of the SDG hierarchy is
equivalent to the existence of ${(r+1)}$-isotonic sections of the
bundle  \gl{(5.26)} for any integer $r\ge 0$.

\bigskip
\section{BRST quantization of the closed N=2 string}

{\bf 6.1\ \ Constraints, ghosts, and pictures}
\smallskip

In Sections 4 and 5 we have described hidden symmetries of the
self-dual gravity equations and the SDG hierarchy generated by the
abelian symmetries.
We now come back to the $N{=}2$ string to present hidden `string'
symmetries, which will be seen to originate from so-called
picture-changing and spectral flow operators naturally occurring
in the BRST approach.
Later on, we will compare those string symmetries with SDG symmetries
and find agreement.

\medskip

The BRST procedure is most convenient for a manifestly $SU(1,1)$
covariant quantization of the $N{=}2$ string~\cite{BL2}.
In the $N{=}2$ superconformal gauge,
the $N{=}2$ supergravity multiplet is reduced to its moduli plus the
super Virasoro constraints
$$
T (z) = - \textstyle{\frac{1}{2}} \partial \bar{y} \cdot \partial y
- \textstyle{\frac{1}{4}} \partial {\psi}^- \cdot \psi^+
- \textstyle{\frac{1}{4}} \partial \psi^+ \cdot {\psi}^-,
$$
$$
G^{+}(z) = \partial \bar{y} \cdot \psi^+ \quad,\qquad
G^{-}(z) = \partial {y} \cdot {\psi}^- \quad,
$$
\be \label{n2currents}
J(z) =  \textstyle{\frac{1}{2}} {\psi}^- \cdot \psi^+
\ee
and their right-moving copy $(\tilde{T},\tilde{G}^\pm,\tilde{J})$.
We have switched to worldsheet light-cone coordinates $(z,\bz)$,
so that $\pa{\equiv}\pa_z$ and $\psi$ is one-component Weyl.
The dot product denotes the $SU(1,1)$ invariant scalar product.

\medskip

A corresponding set of ghost/antighost pairs is introduced,
\be
c,b \qquad \g^\pm,\b^\mp \qquad c',b' \qquad {\rm and} \qquad
\tilde{c},\tilde{b} \qquad \tilde{\g}^\pm,\tilde{\b}^\mp \qquad
\tilde{c}',\tilde{b}' \quad,
\ee
in terms of which the BRST operator
$Q=\oint (j(z)+\tilde{j}(\bar{z}))$ reads
\bea
j\ &=&\  cT+\g^{+} G^{-}+\g^{-} G^{+}+ c'J \nonumber \\[.7ex]
         &&+\,c\pa cb+c\pa c'b'-4\g^{+}\g^{-}b
           +2(\pa\g^{-}\g^{+}{-}\pa\g^{+}\g^{-})b'
           +c'(\g^{+}\b^{-}{-}\g^{-}\b^{+}) \nonumber \\[.7ex]
         &&\,+\frac{3}{4}\pa c(\g^{+}\b^{-}{+}\g^{-}\b^{+})
           -\frac{3}{4}c(\pa\g^{+}\b^{-}{+}\pa\g^{-}\b^{+})
           +\frac{1}{4}c(\g^{+}\pa\b^{-}{+}\g^{-}\pa\b^{+}) \quad.
\eea

\medskip

The canonical commutation resp. anticommutation relations for the
matter and ghost fields are realized on a free field Fock space,
which factorizes into a left- and a right-moving part.
The left-moving (NS) vacuum is defined by
$$
\a_n^A|0\rangle = 0 = \bar{\alpha}_n^\ba|0\rangle
\quad{\rm for}\quad n{\ge} 0 \quad,\qquad
\j_r^{+A}|0\rangle = 0 = \j_r^{-\ba}|0\rangle
\quad{\rm for}\quad r{\ge}{+}\textstyle{\frac12} \quad,
$$
$$
c_n|0\rangle = 0 \quad{\rm for}\quad n{\ge}{+}2 \quad,\qquad
\g_r^\pm|0\rangle = 0 \quad{\rm for}\quad r{\ge}{+}\textstyle{\frac32}
\quad,\qquad c'_n|0\rangle = 0 \quad{\rm for}\quad n{\ge}{+}1\quad,
$$
\be
b_n|0\rangle = 0 \quad{\rm for}\quad n{\ge}{-}1 \quad,\qquad
\b_r^\mp|0\rangle = 0 \quad{\rm for}\quad r{\ge}{-}\textstyle{\frac12}
\quad,\qquad b'_n|0\rangle = 0 \quad{\rm for}\quad n{\ge} 0 \quad.
\ee
Its turns out, however, that the computation of string amplitudes
requires a $\Z^2$-fold copy of our (left-moving NS) Fock space,
labelled by so-called pictures charges $(\pi_+,\pi_-)$.
A second set $(\tilde{\pi}_+,\tilde{\pi}_-)$ of picture charges
occurs for the right-moving Fock space.
The vacuum $|\pi_+,\pi_-\rangle$ of the left-moving $(\pi_+,\pi_-)$
Fock space differs from the above $|0\rangle\equiv|{-}1,{-}1\rangle$
by the relations
\be
\g_r^\pm |\pi_+,\pi_-\rangle\ =\ 0
\quad{\rm for}\quad r\ge\textstyle{\frac12{-}}\pi_\pm
\qquad,\qquad
\b_r^\mp |\pi_+,\pi_-\rangle\ =\ 0
\quad{\rm for}\quad r\ge\textstyle{\frac12}{+}\pi_\pm
\quad.
\ee
Due to the commuting nature of the $(\g,\b)$ ghosts,
different vacua $|\pi_+,\pi_-\rangle$ (and the Fock spaces
built over them) are formally connected only through the action
of {\it distributions\/} like $\d(\g_r^\pm)$ in ghost/antighost modes
or by extending the Fock space via bosonization~\cite{FMS,Ver}.
Hence, different pictures
are isomorphic but not identical. Unless stated otherwise, normal ordering
is understood with respect to $|0\rangle=|{-}1,{-}1\rangle$.
It will turn out that the picture degeneracy is redundant
for the enumeration of physical string states
but crucial for the understanding of the global symmetry structure.

\medskip

It is useful to introduce picture operators $\Pi_\pm$ which measure
the picture charges,
\be
\Pi_+ |\pi_+,\pi_-\rangle\ =\ \pi_+ |\pi_+,\pi_-\rangle \qquad,\qquad
\Pi_- |\pi_+,\pi_-\rangle\ =\ \pi_- |\pi_+,\pi_-\rangle \quad,
\ee
and commute with $Q$.
Another important quantum number is the ghost number~$g=g_L{+}g_R$.
We define the left-moving (chiral) ghost number $g_L$ as the eigenvalue of
\be
G_L\ =\ -\oint \Bigl[ bc + b'c' + \b^+\g^- + \b^-\g^+ \Bigr] - \Pi_+ - \Pi_-
\ee
so that $G_L|\pi_+,\pi_-\rangle=0$, and likewise for $G_R$.
As usual, $Q$ carries $g{=}1$, and $G_L$, $G_R$ commute with $\Pi_\pm$.

\medskip

An additional feature of $N{=}2$ strings is the presence of spectral flow,
which may be interpreted as an isometry of the Maxwell moduli space.
Its effect on an asymptotic string state is a continuous shift in the
monodromies (around the closed string)
of all worldsheet spinors according to their Maxwell charge, thereby
interpolating between NS and R sectors.
Spectral flow acts independently on left- and right-movers.
As it leaves the sum $\pi\equiv\pi_++\pi_-$ invariant
but continuously changes the difference $\D\equiv\pi_+-\pi_-$,
it is not compact but (for $\D{=}2$) relates different (NS) pictures,
\be
S:\quad|\pi_+,\pi_-\rangle\ \longrightarrow\ |\pi_+{+}1,\pi_-{-}1\rangle\quad.
\ee
Actually, this action corresponds to a singular gauge transformation creating
a Maxwell instanton at the puncture and may be used to relate string
amplitude contributions from different instanton sectors~\cite{KL,Le}.
For a given worldsheet topology $(J,M)$, however, the spectral flow isometry
does not alter the amplitude~$A^{(n)}_{J,M}$ since in the sum over all
external string legs the picture shifts and the instanton charges must
cancel out, according to the selection rule
\be
\sum_{i=1}^n (\pi_\pm)_i \ =\ \sum_{i=1}^n (\tilde{\pi}_\pm)_i\ =\
\sfrac12(J\pm M)-n\ =\ 2(\#{\rm handles})-2\pm\sfrac12 M \quad.
\ee
Hence, it suffices to restrict oneself to NS sectors
(i.e. integral pictures) only, as we do in this paper.

\bigskip
\bigskip\noindent
{\bf 6.2\ \ Cohomology for physical states}
\smallskip

It is well known that physical states in gauge theories can be identified
as BRST cohomology classes of a certain ghost number. For the closed $N{=}2$
string, the relevant cohomology is subject to the subsidiary conditions
\be \label{semi}
(b_0-\tilde{b}_0)|\psi\rangle\ =\ 0 \qquad {\rm and} \qquad
b'_0|\psi\rangle\ =\ 0\ =\ \tilde{b}'_0|\psi\rangle
\ee
and may be termed {\it $b$-semirelative\/} and
{\it $b'$-relative\/}.
The BRST cohomology is graded by
\begin{itemize}
\item ghost number $g\in\Z$
\item picture numbers $(\pi_+,\pi_-)\in\Z^2$
\item spacetime momentum $(k^A,\bar{k}^{\bar{A}})$ or $k^{\a\ad}\in\R^{2,2}$
\end{itemize}
so that one may restrict the analysis to Fock states of a given ghost number,
built on picture vacua $|\pi_+,\pi_-;k\rangle$ dressed with momentum.
It is important to distinguish the exceptional case of $k=0$ from the
generic case $(k\ne0)$ which contains the propagating modes.

\medskip

It has been shown~\cite{JL} that the generic BRST cohomology is non-empty
only for
\begin{itemize}
\item $g=2$ and $g=3$
\item any value for $(\pi_+,\pi_-)$
\item any {\it lightlike\/} momentum, $\bar{k}\cdot k=0$
\end{itemize}
and is one-dimensional in each such case.
Since the $g{=}3$ class is merely conjugate to the $g{=}2$ class under
the Fock space scalar product, it does not represent a second physical state.
Alternatively, we may eliminate it by passing to the {\it $b$-relative\/}
cohomology, which imposes
\be
(b_0+\tilde{b}_0)|\psi\rangle\ =\ 0
\ee
on top of the conditions~\gl{semi}.
Since the relative Fock space factorizes into left- and right-movers,
the K\"unneth theorem tells us that the ($b$- and $b'$-)relative BRST
cohomology also factorizes in the same way.
For this reason, it suffices to analyze the relative {\it chiral\/}
BRST cohomology which is graded by $(g_{ch},\pi_+,\pi_-,k)$
and (for lightlike $k{\neq}0$) nonzero only at $g_{ch}{=}1$.\footnote{
Here, $g_{ch}$ denotes $g_L$ or $g_R$, as usual.}
In fact, the physical states (at $g{=}2$)
are just products of left-moving and right-moving parts,
with chiral ghost numbers $g_L{=}g_R{=}1$.

\medskip

Moreover, for $k\ne0$ one may construct {\it picture-raising\/} and
{\it picture-lowering\/} operators which commute with $Q$ and do not carry
ghost number or momentum~\cite{BZ}.
Together with spectral flow, these operators may
therefore be used to define an equivalence relation among all pictures.
This projection of the BRST cohomology leaves us with a single, massless,
physical mode, in accordance with the result of Ooguri and Vafa~\cite{OV}.
It is the string excitation corresponding to the self-dual deformation~$\phi$
of the metric background or to the prepotential $\Psi$ for a self-dual
Riemann tensor.
Yet, the discussion of the global symmetries will provide a justification
for distinguishing different pictures, since different picture-representatives
of the physical mode will be connected with shifts~$\d\phi$ resp. $\d\Psi$.
For a related interpretation of the picture phenomenon, see~\cite{DL1,DL2}.

\bigskip
\bigskip\noindent
{\bf 6.3\ \ Cohomology for global symmetries}
\smallskip

It is perhaps less well known that the exceptional (zero-momentum)
BRST cohomology at ghost number one
harbors all unbroken global symmetry charges of the theory.
The argument goes as follows~\cite{WZ}:

\medskip

A current with components $J_z$ and $J_{\bar z}$ in a two dimensional theory
(complex coordinates $z$ and $\bz$) is conserved, i.e.
satisfies $\bar{\partial} J_z + \partial J_{\bar z} = 0$, when the one-form
\be
\Omega^{(1)}\ =\  J_z dz - J_{\bar z} d\bar{z}
\ee
is closed. The corresponding charge
\be
{\cal A}\ =\ \oint_{\cal C} \Omega^{(1)}
\ee
is conserved if it has the same value for contours $\cal C$ and
${\cal C}'$ that are homologous,
i.e. are the boundaries of some surface $\Sigma$,
$\partial\Sigma = {\cal C}{-}{\cal C}'$. Current conservation implies charge
conservation by Stokes' Theorem.
In BRST quantization these relations are required to  hold  only up
to BRST commutators. Current conservation then reads
\be
d \Omega^{(1)}\ =\ \big[ Q , \Omega^{(2)} \big\}
\ee
for some two form $\Omega^{(2)}= \Omega^{(2)}_{z\bar z} dz \wedge d\bar{z}$.
BRST invariance of the charge requires
$[ Q, \Omega^{(1)} \} = d \Omega^{(0)}$.
Applying $Q$ to this relation implies that $ [ Q, \Omega^{(0)} \}$
is a constant. Furthermore, this constant must vanish, since otherwise
the unit operator were BRST trivial. Summarizing, we
have the descent equations
\bea \label{descent}
\big[ Q, \Omega^{(0)} \big\}\ &=&\ 0 \quad, \nonumber \\[1ex]
\big[ Q, \Omega^{(1)} \big\}\ &=&\ d \Omega^{(0)} \quad, \nonumber \\[1ex]
\big[ Q, \Omega^{(2)} \big\}\ &=&\ d \Omega^{(1)} \quad.
\eea
Moreover, the whole formalism is unaffected by the replacements
\bea
\Omega^{(0)}\ &\rightarrow&\ \Omega^{(0)} +  \big[ Q, \alpha \big\} \quad,
\nonumber \\[1ex]
\Omega^{(1)}\ &\rightarrow&\ \Omega^{(1)} + d\alpha \quad,
\eea
where $\alpha$ is some form of degree zero.
The equations for $\Omega^{(0)}$ are precisely the defining relations for the
BRST cohomology (on operators instead of Fock states).
Taking as $\Omega^{(0)}$ some cohomology class of ghost number~$g$
one can then construct the forms $\Omega^{(1)}$ and $\Omega^{(2)}$
of ghost numbers $g-1$ and $g-2$. It is  most natural to choose
$\Omega^{(0)}$ to have ghost number one. This results in a charge of ghost
number zero that can map physical states to physical states.

\medskip

As a simple example let us consider target space translations in bosonic
string theory. The only ghost number zero chiral cohomology class is the
unit operator, which we may take as the right-moving piece of the closed
string
cohomology. As left-moving piece we must take a ghost number one state that
also has vanishing momentum -- the only candidate is $c \partial x^{\mu}$.
The forms $\Omega^{(i)}$  take the form (suppressing the right-moving unit
operator)
\be
\Omega^{(0)}\ =\ c\,\partial x^{\mu} \quad, \qquad
\Omega^{(1)}\ =\ \partial x^{\mu} dz \quad, \qquad
\Omega^{(2)}\ =\ 0 \quad.
\ee
The charge is just the center-of-mass momentum operator
$p^{\mu} = \oint\!\frac{dz}{2\pi}\,\partial x^{\mu}$.

\medskip

To ensure that the charges $\cal{A}$ generate symmetries on physical states,
one must demand not only that they commute with $Q$ but also that their action
is compatible with our subsidiary conditions,
i.e.
\be
[ {\cal{A}}, b_0  \}\ =\ 0\ =\ [ {\cal{A}}, b'_0 \} \qquad
{\rm and} \qquad
[ {\cal{A}}, \tilde{b}_0 \}\ =\ 0\ =\ [ {\cal{A}}, \tilde{b}'_0 \} \quad.
\ee
Since the ($b$- and $b'$-)relative cohomology factorizes into chiral parts,
we need only compute the exceptional relative {\it chiral\/} cohomology
$H_{rel}^{ch}$. Poincar\'e duality identifies the cohomologies at
\be
(g_{ch},\pi)\ \longleftrightarrow\ (2{-}g_{ch} , -4-\pi) \quad,
\ee
and spectral flow equates cohomologies differing only in the value of~$\D$.
Thus, we may restrict ourselves to $\pi{\ge}-2$ and ignore~$\D$ and
display the current knowledge in the following table~\cite{JL}:\footnote{
Note that there is one error in the corresponding table of \cite{JL}.}

\vspace{0.2cm}
\begin{center}
\begin{tabular}{|r|ccccc|}
\hline
$g_{ch}{\scriptstyle\searrow}$ & & \multicolumn{3}{c}{dim $H_{rel}^{ch}$} & \\
$\pi\downarrow$ & $<0$ & 0 & 1 & 2 & $>2$ \\
\hline
$-2$    & 0 & 0 & 1 & 0 & 0 \\
$-1$    & 0 & 0 & 2 & 2 & 2 \\
$ 0$    & 0 & 1 & 4 & $\ge4$ & $\ge4$ \\
$+1$    & 0 & 2 & $\ge8$ & & \\
$\pi>1$ & 0 & $\ge\pi{+}1$ & $\ge4(\pi{+}1)$ & \phantom{$\ge\pi{+}1$} & \\
\hline
\end{tabular}
\end{center}
\vspace{0.1cm}
Since the multiplication of cohomology classes preserves the ghost number
and picture gradings, the ghost-number zero chiral cohomology by itself forms
a ring, the so-called {\it ground ring\/} $H_{rel}^{ch}(g{=}0)$~\cite{Wi}.
This ground ring is rather important since in some sense it generalizes
the unit operator, and it reveals much about the global symmetries.
It is therefore quite remarkable that the ground ring of the $N{=}2$ string
contains an increasing number of elements in higher pictures.

\medskip

For convenience we change the picture labels from $(\pi_+,\pi_-)$
to ``spin'' labels $(j,m)$ via
\bea
\pi_+\ =\ j+m \qquad {\rm and} & \qquad\, \pi_-\ =\ j-m
\qquad & {\rm on\ operators} \nonumber \\
\pi_+{+}1\ =\ j+m \qquad {\rm and} & \quad \pi_-{+}1\ =\ j-m
\qquad & {\rm on\ states}
\eea
and likewise for the right movers.
The picture offset of the canonical ground state $|-1,-1\rangle$
is responsible for the distinction.
The BRST cohomology possesses a natural multiplication rule~\cite{Wi}
which we denote by a dot.
The (left-moving) ground ring is spanned by the basis elements
\be
{\cal{O}}_{j,m,n}\ =\ (X_+)^{j+m} \cdot (X_-)^{j-m} \cdot H^n
\ee
where the labels range over
\be
j=0,\sfrac12,1,\sfrac32,\ldots \qquad{\rm and}\qquad
m,n{-}m=-j,-j{+}1,\ldots,+j \quad.
\ee
We have introduced the picture-raising operators~\cite{BKL}~\footnote{
Here, $\Theta$ denotes the Heaviside step function.
Note that $X_\pm$ differ from the picture-raisers mentioned at the
end of the previous subsection.}
\be
X_\pm\ =\ c\,\pa\Theta(\b^\pm) + \d(\b^\pm)\,(G^\pm - 4\g^\pm b
\mp 4\pa\g^\pm b' \mp 2\g^\pm\pa b')
\ee
as well as picture-neutral formal operator~$H$ via
\be
X_+ \cdot H\ =\ X_- \cdot S \qquad{\rm and}\qquad
X_- \cdot H^{-1}\ =\ X_+ \cdot S^{-1} \quad,
\ee
composed from the spectral flow operator
\be
S\ =\  \textstyle{\frac14}\,\d(\g^+)\,\d(\b^+)\,
\e_{AB}\j^{+A}\j^{+B}\,(1{+}cb')
\ee
and its inverse
\be
S^{-1}\ =\ -\textstyle{\frac14}\,\d(\b^-)\,\d(\g^-)\,
\e_{\ba\bb}\j^{-\ba}\j^{-\bb}\,(1{-}cb')
\quad.
\ee
The $\D{=}0$ representatives obtain for $m{=}0$ and $j{=}\pi/2$.
It is important to note that, while $S$ can be inverted, $X_\pm$ do
not have inverses~\cite{BKL}. This fact limits the range of $n$
because each factor of $H$ ($H^{-1}$) must be matched by a factor of
$X_+$ ($X_-$) to make sense.

\medskip

Let us define yet another combination of picture-raising and spectral
flow operators, namely
\be
Y_-\ =\ X_+ \cdot S^{-1}\ =\ X_- \cdot H^{-1} \quad,
\ee
which allows us to pull out a non-negative power $\ell=j-m+n$ of~$H$ in
\be
{\cal{O}}_{j,m,n}\ =\
(X_+)^{j+m} \cdot (Y_-)^{j-m} \cdot H^{\ell}\ =:\
{\cal{O}}^{\ell}_{j,m} \qquad{\rm with}\quad \ell=0,1,\ldots,2j \quad.
\ee
The chiral ground ring operators ${\cal{O}}^{\ell}_{j,m}$ form an
infinite abelian algebra under the cohomology product,
\be
{\cal{O}}^{\ell}_{j,m} \cdot {\cal{O}}^{\ell'}_{j',m'}\ =\
{\cal{O}}^{\ell+\ell'}_{j+j',m+m'} \quad.
\ee
The picture operators $\Pi_\pm$ act as derivations on the ground ring,
\be
[\,\Pi_\pm\,,\,{\cal{O}}^{\ell}_{j,m}\,]\ =\ (j\pm m)\,{\cal{O}}^{\ell}_{j,m}
\quad,
\ee
and represent only the tip of an iceberg,
namely an infinite-dimensional algebra
of derivations spanned by
\be
{\cal{D}}^{+,\ell}_{j,m}\ =\ {\cal{O}}^{\ell}_{j,m}\cdot(\Pi_+{+}1)
\qquad{\rm and}\qquad
{\cal{D}}^{-,\ell}_{j,m}\ =\ {\cal{O}}^{\ell}_{j,m}\cdot(\Pi_-{+}1) \quad.
\ee
The remaining commutators are
\bea
[\,{\cal{D}}^{\pm,\ell'}_{j',m'}\,,\,{\cal{O}}^{\ell}_{j,m}\,]\ &=&\
(j{\pm}m)\,{\cal{O}}^{\ell+\ell'}_{j+j',m+m'} \quad,\nonumber \\[1ex]
[\,{\cal{D}}^{\pm,\ell'}_{j',m'}\,,\,{\cal{D}}^{+,\ell}_{j,m}\,]\ &=&\
(j{\pm}m)\,{\cal{D}}^{+,\ell+\ell'}_{j+j',m+m'} -
(j'{+}m') \,{\cal{D}}^{\pm,\ell+\ell'}_{j+j',m+m'} \quad, \nonumber \\[1ex]
[\,{\cal{D}}^{\pm,\ell'}_{j',m'}\,,\,{\cal{D}}^{-,\ell}_{j,m}\,]\ &=&\
(j{\pm}m)\,{\cal{D}}^{-,\ell+\ell'}_{j+j',m+m'} -
(j'{-}m') \,{\cal{D}}^{\pm,\ell+\ell'}_{j+j',m+m'} \quad.
\eea

\medskip

The algebra of the ground ring and its derivations can be written
more concisely in terms of polynomials in two variables $(x,y)$
and vector fields on the $(x,y)$ plane.
We first remark that the neutral operator $H$ can be expressed in terms
of momentum operators $p^A$ and $\bar{p}^\ba$, so that on momentum eigenstates
it may be replaced by its eigenvalue, a function~$h(k)$.
We also like to point out that the picture-raisers $X_\pm$ do not carry
any spacetime quantum numbers whereas the spectral flow~$S$ (and therefore~$H$)
transform non-trivially under ``Lorentz'' transformations~\cite{Le}.
To keep the spacetime properties manifest it is thus convenient to define
\be
x\ :=\ X_+\ =\ {\cal{O}}^0_{\frac12,+\frac12} \qquad{\rm and}\qquad
y\ :=\ X_-\ =\ Y_- \cdot H\ =\ {\cal{O}}^0_{\frac12,-\frac12} \cdot H
\ee
as generators of a ring of polynomials in two variables and to allow for
multiplication with arbitrary powers of~$h$.
In these variables,
\be
{\cal{O}}^{\ell}_{j,m}\ =\
(X_+)^{j+m} \cdot (X_-)^{j-m} \cdot H^{-(j-m)+\ell}\ =\
x^{j+m}\,y^{j-m}\,h^{-(j-m)+\ell} \quad,
\ee
and the elementary derivations become
\be
\Pi_+{+}1 = x\,\pa_x \qquad {\rm and} \qquad \Pi_-{+}1 = y\,\pa_y \quad,
\ee
representing the abelian part of the huge algebra of derivations
${\cal{D}}^{\pm,\ell}_{j,m}$ by vector fields on the $(x,y)$ plane,
with elements
\be
p_x(x,y,h)\,x\,\pa_x\ +\ p_y(x,y,h)\,y\,\pa_y
\ee
containing functions $p_x$ and $p_y$ polynomial in $x$ and $y$.
Note that the translation generators $\pa_x$ and $\pa_y$ are missing.

\medskip

For the complete story, one must combine left- and right-movers.
As stated above, symmetry charges carry total ghost number one.
Consulting our table, the only way to build a $g{=}1$ representative is to
combine a $g_R{=}0$ class with a $g_L{=}1$ class, or vice versa.
Only counting the first possibility, we arrive at least at
$4(\pi{+}1)(\tilde{\pi}+1)$ cohomology classes for non-negative values
of $\pi$ and $\tilde{\pi}$, while the $\pi{=}-1,-2$ sectors are empty.

\medskip

Since the canonical $g_L{=}1$ representatives are just the products
$P^A\cdot{\cal{O}}^{\ell}_{j,m}$ and $\bar{P}^\ba\cdot{\cal{O}}^{\ell}_{j,m}$,
with
\be
-i P^A\ =\ c\,\p y^A - 2\g^-\j^{+A}
\qquad{\rm and}\qquad
-i \bar{P}^\ba\ =\ c\,\p \by^\ba - 2\g^+\j^{-\ba} \quad,
\ee
we obtain the $4(2j{+}1)(2\tj{+}1)$ operators
\be
\Omega^{A;\ell;\tl\ (0)}_{j,m;\tj,\tm}\ =\
P^A\cdot{\cal{O}}^{\ell}_{j,m}\,\tilde{\cal{O}}^{\tl}_{\tj,\tm}
\qquad{\rm and}\qquad
\Omega^{\ba;\ell;\tl\ (0)}_{j,m;\tj,\tm}\ =\
\bar{P}^\ba\cdot{\cal{O}}^{\ell}_{j,m}\,\tilde{\cal{O}}^{\tl}_{\tj,\tm}
\ee
in the $(j,m;\tj,\tm)$ sector with $g{=}1$.

\medskip

In order to find the symmetry charges $\cal{A}$,
we have to insert our $g{=}1$ zero-forms
$\Omega^{(0)\ A;\ell;\tl}_{j,m;\tj,\tm}$ and
$\Omega^{(0)\ \ba;\ell;\tl}_{j,m;\tj,\tm}$
into the descent equations~\gl{descent}
and work out the corresponding one-forms
$\Omega^{(1)\ A;\ell;\tl}_{j,m;\tj,\tm}$ and
$\Omega^{(1)\ \ba;\ell;\tl}_{j,m;\tj,\tm}$,
to be integrated around some contour.
This computation has been performed in ref.~\cite{JLP}, with the result
\bea
{\cal{A}}^{A;\ell;\tl}_{j,m;\tj,\tm}\ &=&\
\oint\!\frac{dz}{2\pi i} \Bigl[ \oint_z\!\frac{dw}{2\pi i}\, b(w)\,
P^A(z)\cdot{\cal{O}}^{\ell}_{j,m}(z)\,
\tilde{\cal{O}}^{\tl}_{\tj,\tm}(\bz) \Bigr] \nonumber \\[1ex]
&&-\,\oint\!\frac{d\bz}{2\pi i} \Bigl[ P^A(z)\cdot{\cal{O}}^{\ell}_{j,m}(z)\,
\oint_\bz\!\frac{d\bw}{2 \pi i}\,\tilde{b}(\bw)\,
\tilde{\cal{O}}^{\tl}_{\tj,\tm}(\bz) \Bigr]
\eea
and likewise for ${\cal{A}}^{\ba;\ell;\tl}_{j,m;\tj,\tm}$.

Combining left- and right-movers,
one arrives also at more general operators
\be
{\cal{B}}^{\pm,\ell;\tl}_{j,m;\tj,\tm}\ =\
{\cal{O}}^{\ell}_{j,m}\cdot{\tilde{\cal{O}}}^{\tl}_{\tj,\tm}
\cdot(\Pi_\pm{+}1)
\qquad{\rm and}\qquad
{\tilde{\cal{B}}}^{\ell;\pm,\tl}_{j,m;\tj,\tm}\ =\
{\cal{O}}^{\ell}_{j,m}\cdot{\tilde{\cal{O}}}^{\tl}_{\tj,\tm}
\cdot(\tilde{\Pi}_\pm{+}1)
\ee
which, together with the charges
${\cal{A}}^{A;\ell;\tl}_{j,m;\tj,\tm}$ and
${\cal{A}}^{\ba;\ell;\tl}_{j,m;\tj,\tm}$,
form an enormous non-abelian algebra.

\bigskip
\section{Symmetry transformations of physical states}

For the discussion of string symmetries, it is revealing not to identify
the physical state representatives in the various picture sectors
but to distinguish the (boosted) picture-vacua
$|\pi_+,\pi_-;\tilde{\pi}_+,\tilde{\pi}_-;k\rangle$
as different states.
Clearly, those states are obtained by applying the ground ring operators
${\cal{O}}^{\ell}_{j,m}$ and $\tilde{\cal{O}}^{\tl}_{\tj,\tm}$
to the canonical representative, $|-1,-1;-1,-1;k\rangle$.
Focusing on the left-moving part, one obtains
\be
{\cal{O}}^{\ell}_{j,m}|-1,-1;k\rangle\ =\
h(k)^{-(j-m)+\ell}\,|j{+}m{-}1,j{-}m{-}1;k\rangle\
\ee
where the phase~\cite{Par,BV}
\be
h(k)\ =\ {k^1}/{\bar{k}^{\bar{2}}}\ =\ {k^2}/{\bar{k}^{\bar{1}}}\ =\
(h(k)^{-1})^*
\ee
is just the eigenvalue of our formal operator~$H$.
Hence, there is an equivalence
\be
|j{+}m{-}1,j{-}m{-}1;k\rangle\ \sim\ x^{j+m}\,y^{j-m}
\ee
modulo powers of $h(k)$.

\medskip

By virtue of this equivalence, the operators
\be
{\cal{D}}^{+,\ell}_{j,m}\ =\
h^{-(j-m)+\ell}\,x^{j+m+1}\,y^{j-m}\,\pa_x
\qquad{\rm and}\qquad
{\cal{D}}^{-,\ell}_{j,m}\ =\
h^{-(j-m)+\ell}\,x^{j+m}\,y^{j-m+1}\,\pa_y
\ee
have a natural action on states as well: they not only shift the quantum
numbers $(\pi_+,\pi_-)$ of a state but also pull out an additional factor
of $\pi_\pm{+}1$ from it, since we have added unity to $\Pi_\pm$
to account for the picture offset of the $j{=}0$ ground state.
Thus, we may alternatively use ${\cal{D}}^{\pm,\ell}_{j,m}$
(and their right-moving images) to generate all physical states
from $|0,-1;k\rangle$ and $|-1,0;k\rangle$ (times their right-moving partners).

\medskip

In order to find the transformation laws of the full physical states,
we have to apply ${\cal{A}}^{A;\ell;\tl}_{j,m;\tj,\tm}$
and ${\cal{B}}^{\pm,\ell;\tl}_{j,m;\tj,\tm}$.
The action of these charges on the physical states factorizes,\footnote{
In these equations, picture and spin labels are unrelated, in order
to distinguish between the quantum numbers of the operator and those
of the state.}
\bea \label{acta}
&& {\cal{A}}^{A;\ell;\tl}_{j,m;\tj,\tm}\,
|\pi_+,\pi_-;\tilde{\pi}_+,\tilde{\pi}_-;k\rangle\ =\
{\cal{O}}^{\ell}_{j,m}\cdot\tilde{\cal{O}}^{\tl}_{\tj,\tm}\cdot p^A\,
|\pi_+,\pi_-;\tilde{\pi}_+,\tilde{\pi}_-;k\rangle  \nonumber \\[1ex]
&& \qquad =\ h(k)^{-(j-m)-(\tj-\tm)+\ell+\tl}\,k^A\,
|\pi_+{+}j{+}m,\pi_-{+}j{-}m;
\tilde{\pi}_+{+}\tj{+}\tm,\tilde{\pi}_-{+}\tj{-}\tm;k\rangle \  ,
\eea
\bea \label{actb}
&& {\cal{B}}^{\pm,\ell;\tl}_{j,m;\tj,\tm}\,
|\pi_+,\pi_-;\tilde{\pi}_+,\tilde{\pi}_-;k\rangle\ =\
{\cal{O}}^{\ell}_{j,m}\cdot\tilde{\cal{O}}^{\tl}_{\tj,\tm}\cdot(\Pi_\pm{+}1)\,
|\pi_+,\pi_-;\tilde{\pi}_+,\tilde{\pi}_-;k\rangle  \nonumber \\[1ex]
&& \qquad =\ h(k)^{-(j-m)-(\tj-\tm)+\ell+\tl}\,(\pi_\pm{+}1)\,
|\pi_+{+}j{+}m,\pi_-{+}j{-}m;
\tilde{\pi}_+{+}\tj{+}\tm,\tilde{\pi}_-{+}\tj{-}\tm;k\rangle \ ,
\eea
and similarly for $A\to\ba$ or ${\cal{B}}\to\tilde{\cal{B}}$.
These transformations constitute an infinity of global symmetries
which are unbroken in the flat Kleinian background.
Their Ward identities constrain the tree-level scattering amplitudes
so severely~\cite{JLP} that all but the three-point function must vanish,
consistent with the direct computations alluded to earlier.

\medskip

To make contact with the abelian symmetries of self-dual gravity (see Sect.5),
it suffices to consider the sub-algebra of symmetry charges
${\cal{A}}^{\ba;0;0}_{j,m;\tj,\tm}$, i.e. putting $\ell{=}0{=}\tl$.
It has the important property that only non-positive powers of the
phase~$h(k)$ appear when acting by these charges on physical states.
Moreover, since the factor $h^{-(j+\tj-m-\tm)}$ in \gl{acta} feels neither
of the differences $j{-}\tj$ and $m{-}\tm$, one can introduce states
\be \label{symstates}
||j,m;k\rangle\rangle\ :=\!
\sum_{{j_1,j_2\atop j_1+j_2=j}} \sum_{{m_1,m_2\atop m_1+m_2=m}}\!
|j_1{+}m_1{-}1,j_1{-}m_1{-}1;j_2{+}m_2{-}1,j_2{-}m_2{-}1;k\rangle
\ee
which are {\it symmetric\/} tensor products of left- and right-moving states.
These states form a $(2j{+}1)$plet for a fixed value of $j$
since $m=-j,-j{+}1,\ldots,j$. We also introduce combinations of charges,
\be
{\cal P}_A^{j-m}\ :=\ \eta_{A\ba}\!
\sum_{{j_1,j_2\atop j_1+j_2=j}} \sum_{{m_1,m_2\atop m_1+m_2=m}}\!
{\cal{A}}^{\ba;0;0}_{j_1,m_1;j_2,m_2}
\ee
corresponding to the symmetric tensor product states~\gl{symstates}.
The action of these charges on the ground state
\be
||0,0;k\rangle\rangle\ =\ |-1,-1;-1,-1;k\rangle
\ee
has the form
\be \label{transstr}
\d_A^{j-m}||0,0;k\rangle\rangle\ :=\
{\cal P}_A^{j-m}||0,0;k\rangle\rangle\ =\
k_A\,h^{-(j-m)}\,||j,m;k\rangle\rangle \quad.
\ee
Note that for $j{=}0$ we find the translations
$\d_A^0||0,0;k\rangle\rangle=k_A||0,0;k\rangle\rangle$ as it should be.

\medskip

Recall that, for the SDG equations, abelian symmetries generated by
recursion from translations
$\pa/\pa y^A=\pa/\pa t_0^A$ ($=k_A$ in the momentum representation) read
\be
\d_A^n\,\phi(y^B,\by^{\bar B},t^B_1,\dots)\ =\
\frac{\pa}{\pa t^A_n}\,\phi(y^B,\by^{\bar B},t^B_1,\ldots)\quad,\qquad
n=0,1,\ldots \ ,
\ee
where $t^A_1,\ldots,t^A_n,\ldots$ parametrize the moduli space of
solutions $\phi$ to the SDG equations (see Sect.5).
If we impose on $\phi$ the condition (5.37), i.e. consider the
{\it finite\/} multiplet of symmetries
\be
\d_A^{j-m}\,\phi\ =\ \frac{\pa}{\pa t^A_{j-m}}\,\phi
\ee
with $j$ fixed and $m=-j,\ldots,j$, then in the momentum
representation we will have
\be \label{transSDG}
\d_A^{j-m}\,\tilde\phi(k)\ =\ p_A^{j-m}\,\tilde\phi(k)
\ee
where $p_A^{j-m}$ are the momenta corresponding to $\pa/\pa t^A_{j-m}$.
Notice that $p_A^{j-m}$ implicitly depends on $\tilde\phi(k)$ which renders
the transformations \gl{transSDG} non-linear in $\tilde\phi(k)$.

\medskip

As was argued in \cite{JLP}, only the linear part of the symmetries
\gl{transSDG} can be seen in first-quantized string theory,
since in this framework all spacetime symmetries are connected to
BRST cohomology. Moreover, in \cite{JLP} it was shown that the
transformations \gl{transSDG} can be rewritten in the form
\be \label{translin}
\d_A^{j-m}\,\tilde\phi(k)\ =\
k_A\,h(k)^{-(j-m)}\,\tilde\phi(k)\ +\ O(\tilde\phi^2) \quad.
\ee
These are just the non-local transformations derived recursively from
the global translations $y^A\to y^A+t^A_0$.
The similarity of \gl{transstr} to \gl{translin} is evident.
Thus, the transformations \gl{transstr} of the string ground state
$||0,0;k\rangle\rangle$ corresponding to the field $\tilde\phi(k)$
precisely reproduce the linear part of the symmetries \gl{transSDG}
of self-dual gravity.

\section{ Conclusion }

This work deepens the identity of closed $N{=}2$ strings
(at tree-level) with self-dual gravity (SDG) on $4D$ manifolds of
signature $(++-\,-)$, by the following results.
After outlining how the SDG equations emerge in different gauges
(Plebanski I and II) from the string dynamics,
we have analyzed in detail the non-local hidden symmetries of the
SDG equations and introduced an infinite hierarchy of equations
(the SDG hierarchy) associated with an infinite number of abelian symmetries.
We have then demonstrated how exactly the same symmetries (in linearized form)
emerge in first-quantized $N{=}2$ closed string theory.

\medskip

Interestingly, the stringy source of those symmetries is a non-trivial
ground ring of ghost number zero operators in the chiral BRST cohomology.
Such a phenomenon is familiar from the non-critial $2D$ string,
where the ground ring was exploited to investigate the global symmetries
of the theory, with the result that there are more discrete states and
associated symmetries in $2D$ string theory than had been recognized
previously~\cite{Wi,WZ}. The authors of ref.~\cite{WZ} have wondered
if their findings ``could be relevant in a realistic string theory
with a macroscopic four-dimensional target space''.
The outcome of this paper answers their question in the affirmative
by providing the first four-dimensional if not yet realistic string
theory with a rich symmetry structure based on an infinite ground ring.

\medskip

More concretely, the symmetry charges are constructed from zero
momentum operators of picture-raising $X_\pm$, picture charge $\Pi_\pm$,
spectral flow $S$, and momentum operators $p^A$, $\bar{p}^\ba$.
The abelian subalgebra generated by $X_\pm$, $S$, and $\h_{A \ba}\,\bar{p}^\ba$
was found to coincide with the abelian symmetries of the SDG equations
produced by the operators $\p_A$ and the recursion operator $\mr$.

\medskip

Our results ascertain that the non-trivial picture structure of the BRST
cohomology is not just an irrelevant technical detail of the BRST approach
but indispensible for a deeper understanding of the theory.
It is, of course, not a simple task to discover the full symmetry group
of a string model. Doing so would roughly correspond to having found a
useful non-perturbative definition of the theory. In this paper we have
worked in the standard first-quantized formalism which is background-dependent
and limits our access to unbroken linear global symmetries.

\medskip

There remain a number of interesting unresolved issues which should
be addressed.
Prominent among them is the detection of the {\it non-abelian\/}
symmetries of the SDG equations in the $N{=}2$ string context.
Such symmetries constitute an affinization of the $W_\infty$ algebra
of volume-preserving $2D$ vector fields. In view of our findings here,
it is tempting to relate them to the non-abelian string symmetries
generated by the operators ${\cal{B}}^{\pm,\ell;\tl}_{j,m;\tj,\tm}$
and $\tilde{\cal{B}}^{\ell;\pm,\tl}_{j,m;\tj,\tm}$ introduced in Sect.6.3.
Another task is the transfer of our results to the {\it open\/} $N{=}2$
string pertaining to the self-dual Yang-Mills equations, in Leznov or in
Yang gauge.
Finally, it would be interesting to find further examples in which
the picture structure yields non-trivial information about a theory,
like it happens for the relative zero-momentum cohomology of the Ramond sector
of the $N{=}1$ string in flat $9{+}1$ dimensional spacetime~\cite{BZ}.
We hope to return to these problems soon.


\vspace{1.5cm}
\noindent
{\large\bf Acknowledgment}

\medskip
\noindent
A.D.P. thanks the Institut f\"ur Theoretische Physik der Universit\"at
Hannover for its hospitality.
The work of A.D.P. was partially supported by the grant RFBR-99-01-01076
and ``Freundeskreis der Universit\"at Hannover e.V.''.
O.L. acknowledges support by DFG under the grant LE-838/5.


\vfill\eject
\noindent
\addcontentsline{toc}{appe}{\medskip {\bf References}\hfill}



\baselineskip=14pt
\begin{thebibliography}{99}

\bibitem{OV} H. Ooguri and C. Vafa,
        {\it Geometry of N=2 strings},
        {\sl Nucl. Phys.} {\bf B361} (1991) 469-518;\\
        {\it Self-duality and N=2 string magic},
        {\sl Mod. Phys. Lett.} {\bf A5} (1990) 1389-1398.

\bibitem{LS} O. Lechtenfeld and W. Siegel,
        {\it N=2 worldsheet instantons yield cubic self-dual Yang-Mills},
        {\sl Phys. Lett.} {\bf B405} (1997) 49-54, hep-th/9704076.

\bibitem{Ple} J.F. Plebanski,
        {\it Some solutions of complex Einstein equations},\\
        {\sl J. Math. Phys.} {\bf 16} (1975) 2395-2402.

\bibitem{Ta} K. Takasaki,
        {\it Symmetries of hyperkahler (or poisson gauge field) hierarchy},\\
        {\sl J. Math. Phys.} {\bf 31} (1990) 1877-1888.

\bibitem{Pa} Q.-H. Park,
        {\it 2D sigma model approach to 4d instantons },\\
        {\sl Int. J. Mod. Phys.} {\bf A7} (1992) 1415-1447.

\bibitem{PBR} A.D. Popov, M. Bordemann and H. R\"omer,\\
        {\it Symmetries, currents and conservation laws of self-dual gravity},\\
        {\sl Phys. Lett.} {\bf B385} (1996) 63-74, hep-th/9606077.

\bibitem{JLP} K. J\"unemann, O. Lechtenfeld and A.D. Popov,
        {\it Non-local symmetries of the closed N=2 string},\\
        {\sl Nucl. Phys.} {\bf B548} (1999) 449-474, hep-th/9901164.

\bibitem{FMS} D. Friedan, E. Martinec and S. Shenker,\\
        {\it Conformal invariance, supersymmetry and string theory},
        {\sl Nucl. Phys.} {\bf B271} (1986) 93-165.

\bibitem{JL} K. J\"unemann and O. Lechtenfeld, \\
        {\it Chiral BRST cohomology of the N=2 string at arbitrary
             ghost and picture number},\\
        {\sl Comm. Math. Phys.} {\bf 203} (1999) 53-69, hep-th/9712182.

\bibitem{BZ} N. Berkovits and B. Zwiebach,
        {\it On the picture dependence of Ramond-Ramond cohomology},\\
        {\sl Nucl. Phys.} {\bf B523} (1998) 311-343, hep-th/9711087.

\bibitem{BS} L. Brink and J.H. Schwarz,
        {\it Local complex supersymmetry in two dimensions},\\
        {\sl Nucl. Phys.} {\bf B121} (1977) 285-295.

\bibitem{BL1} J. Bischoff and O. Lechtenfeld,
        {\it Restoring reality for the self-dual N=2 string},\\
        {\sl Phys. Lett.} {\bf B390} (1997) 153, hep-th/9608196.

\bibitem{Bi} J. Bie\'nkowska,
        {\it The generalized no-ghost theorem for N=2 SUSY critical strings},\\
        {\sl Phys. Lett.} {\bf B281} (1992) 59-66, hep-th/9111047.

\bibitem{Hipp} R. Hippmann,
        {\it Tree-Level Amplituden des N=2 String},
        diploma thesis ITP Hannover, \\ September 1997,
        http://www.itp.uni-hannover.de/\~{}lechtenf/Theses/hippmann.ps.

\bibitem{BGPPR}J. Barrett, G.W. Gibbons, M.J. Perry, C.N. Pope and P. Ruback,\\
        {\it Kleinian geometry and the N=2 superstring},\\
        {\sl Int. J. Mod. Phys.} {\bf A9} (1994) 1457-1494, hep-th/9302073.

\bibitem{AFM} L. Alvarez-Gaum\'e, D.Z. Friedman and S. Mukhi,\\
        {\it The background field method and the ultraviolet structure
             of the supersymmetric nonlinear sigma model},
        {\sl Annals Phys.} {\bf 134} (1981) 85-135.

\bibitem{Pe} R. Penrose,
        {\it Non-linear gravitons and curved twistor theory},\\
        {\sl Gen. Rel. Grav.} {\bf 7} (1976) 31-52.

\bibitem{AHS} M.F. Atiyah, N.J. Hitchin and I.M. Singer,\\
        {\it Self-duality in four-dimensional Riemannian geometry},\\
        {\sl Proc. R. Soc. Lond.} {\bf A362} (1978) 425-461.

\bibitem{TW} K.P. Tod and R.S. Ward,
        {\it Self-dual metrics with self-dual Killing vectors},\\
        {\sl Proc. R. Soc. Lond.} {\bf A368} (1979) 411-427.

\bibitem{Wa} R.S.Ward,
        {\it Self-dual space-time with cosmological constant},\\
        {\sl Commun. Math. Phys.} {\bf 78} (1980) 1-17.

\bibitem{HKLR} N.J. Hitchin, A. Karlhede, U. Lindstr\"om and M. Ro\v{c}ek,\\
        {\it Hyperk\"ahler metrics and supersymmetry},
        {\sl Commun. Math. Phys.} {\bf 108} (1987) 535-589.

\bibitem{MW} L.J. Mason and N.M.J. Woodhouse,
        {\it Integrability, self-duality and twistor theory},\\
        Oxford University Press, Oxford, 1996.

\bibitem{Wo} N.M.J. Woodhouse,
        {\it Contour integrals for the ultrahyperbolic wave equation},\\
        {\sl Proc. R. Soc. Lond.} {\bf A438} (1992) 197-206.

\bibitem{Po} A.D. Popov,
        {\it Holomorphic Chern-Simons-Witten theory:
        from $2D$ to $4D$ conformal field theory},\\
        {\sl Nucl. Phys.} {\bf B550} (1999) 589-621, hep-th/9806239.

\bibitem{Hi} N.J. Hitchin,
        {\it Linear field equations on self-dual spaces},\\
        {\sl Proc. R. Soc. Lond.} {\bf A370} (1980) 173-191.

\bibitem{Gi} S.G. Gindikin,
        in: {\it Twistor geometry and non-linear systems},\\
        H.D. Doebner and T. Weber (eds.), Lect. Notes Math. Vol. 970,
        Springer, Berlin, 1982, p.2.

\bibitem{Sa} M. Sato and Y. Sato,\\
        {\it Soliton equations as dynamical systems on infinite
         dimensional Grassmann manifolds},\\
        Lecture Notes in Num. Appl. Anal. {\bf 5} (1982) 259-271.

\bibitem{DJKM} E. Date, M. Jimbo, M. Kashiwara and T. Miwa,
        {\it Transformation groups for soliton equations},\\
        in: {\it Nonlinear integrable systems --- classical
         theory and quantum theory},\\
        M. Jimbo and T. Miwa (eds.),
        World Scientific, Singapore, 1983, pp.39-120.

\bibitem{SW} G. Segal and G. Wilson,
        {\it Loop groups and equations of KdV type},\\
        {\sl Publ. Math. IHES} {\bf 61} (1985) 5-65.

\bibitem{Ne} A.C. Newell,
        {\it Solitons in mathematics and physics}, SIAM, Philadelphia, 1985.

\bibitem{AG} D.V. Alekseevsky and M.M. Graev,\\
        {\it Grassmann and hyperK\"ahler structures on some spaces
         of sections of holomorphic bundles},\\
        in: {\it Manifold and Geometry}, Sympos. Math. 36,
        N. Hitchin, P. de Bartolomeis, F. Tricerri and E. Vesentini (eds.),
        Cambridge University Press, Cambridge, 1996, pp.1-19.

\bibitem{Ko} K. Kodaira,
        {\it A theorem of completeness of characteristic systems for
         analytic families of compact submanifolds of complex manifolds},
        {\sl Ann. Math.} {\bf 84} (1962) 146-162.

\bibitem{BL2} J. Bischoff and O. Lechtenfeld,
        {\it Path-integral quantization of the (2,2) string},\\
        {\sl Int. J. Mod. Phys.} {\bf A12} (1997) 4933-4972, hep-th/9612218.

\bibitem{Ver} E. Verlinde and H. Verlinde,
        {\it Multiloop calculations in covariant superstring theory},\\
        {\sl Phys. Lett.} {\bf B192} (1987) 95-110.

\bibitem{KL} S.V. Ketov and O. Lechtenfeld,
        {\it The string measure and spectral flow of critical N=2 strings},\\
        {\sl Phys. Lett.} {\bf B353} (1995) 463-470, hep-th/9503232.

\bibitem{Le} O. Lechtenfeld,
        {\it Integration measure and spectral flow
         in the critical N=2 string},\\
        {\sl Nucl. Phys.} (Proc. Suppl.) {\bf B49} (1996) 51-56, hep-th/9512189.

\bibitem{DL1} C. Devchand and O. Lechtenfeld,
        {\it Extended self-dual Yang-Mills from the N=2 string},\\
        {\sl Nucl. Phys.} {\bf B516} (1998) 255-272, hep-th/9712043.

\bibitem{DL2} C. Devchand and O. Lechtenfeld,
        {\it String induced Yang-Mills coupling to self-dual gravity},\\
        {\sl Nucl. Phys.} {\bf B539} (1999) 300-328, hep-th/9808053.

\bibitem{WZ} E. Witten and B. Zwiebach,
        {\it Algebraic structures and differential geometry
         in 2-d string theory},\\
        {\sl Nucl. Phys.} {\bf B 377} (1992) 55-112, hep-th/9201056.

\bibitem{Wi} E. Witten,
        {\it Ground ring of two-dimensional string theory},\\
        {\sl Nucl. Phys.} {\bf B 373} (1992) 187-213, hep-th/9108004.

\bibitem{BKL} J. Bischoff, S.V. Ketov and O. Lechtenfeld,\\
        {\it The GSO projection, BRST cohomology and picture-changing
             in N=2 string theory},\\
        {\sl Nucl. Phys.} {\bf B438} (1995) 373-410, hep-th/9406101.

\bibitem{Par} A. Parkes, \\
        {\it On N=2 strings and classical scattering solutions of
         self-dual Yang-Mills in (2,2) space-time},\\
        {\sl Nucl. Phys.} {\bf B 376} (1992) 279-296, hep-th/9110075.

\bibitem{BV} N. Berkovits and C. Vafa,
        {\it N=4 topological strings},\\
        {\sl Nucl. Phys.} {\bf B 433} (1995) 123-180, hep-th/9407190.


\end{thebibliography}
\end{document}